%% file: main.tex
\documentclass[journal,10pt]{IEEEtran}
\usepackage{cite}
\usepackage{amsmath,amssymb,amsfonts}
\usepackage{algorithmic}
\usepackage{lipsum,adjustbox}
\usepackage{verbatim}
\usepackage{graphics}
\usepackage{stackengine}
\usepackage{pgfplots}
\pgfplotsset{width=7cm,compat=1.8}
\usepackage{multirow}
\usepackage{soul}
\usepackage{amssymb}
\usepackage{amsmath}
\usepackage{algorithm}
\usepackage{algorithmic}
\usepackage{graphicx}
\usepackage{subcaption}
\usepackage{textcomp}
\usepackage{booktabs}
\usepackage{stfloats}
\usepackage{scalefnt}
\usepackage{mathrsfs}       
\usepackage[nolist]{acronym}
\input{commands}
\input{acronyms.tex}
\usepackage[utf8]{inputenc}
\usepackage[english]{babel}
\usepackage{amsthm}
\usepackage{commath}

\begin{document}

\title{CNN aided Weighted Interpolation for Channel Estimation in Vehicular Communications}

\author{Abdul~Karim~Gizzini,~Marwa~Chafii,~Ahmad~Nimr,~Raed~M.~Shubair,~and~Gerhard~Fettweis


\thanks{

Authors acknowledge the CY INEX for the support of the project through the ASIA Chair of Excellence Grant (PIA/ANR-16-IDEX-0008).

Abdul Karim Gizzini is with ETIS, UMR8051, CY Cergy Paris Université, ENSEA, CNRS, France (e-mail: abdulkarim.gizzini@ensea.fr).

 Marwa Chafii and Raed M. Shubair are with the Department of Electrical and Computer Engineering, New York University (NYU), Abu Dhabi 129188, UAE (e-mail: \{marwa.chafii, raed.shubair\}@nyu.edu).

Ahmad Nimr and Gerhard Fettweis are with the Vodafone Chair Mobile Communication Systems, Technische Universitat Dresden, Germany (e-mail: \{ahmad.nimr, gerhard.fettweis\}@tu-dresden.de).

}
}
\markboth{} 
{}
\maketitle
\input{abstract}
\IEEEpeerreviewmaketitle
\renewcommand{\figurename}{Fig.}
\input{introduction}
\input{system_model}

\input{SoA_estimators}
\input{Proposed_2DWI}

\input{CNN}

\input{Simulation_Results_Updated}
\input{complexity}

\input{conclusions}
\ifCLASSOPTIONcaptionsoff
  \newpage
\fi
\bibliographystyle{IEEEtran}
\bibliography{ref}


\begin{IEEEbiography}[{\includegraphics[width=1in,height=1.25in,clip,keepaspectratio]{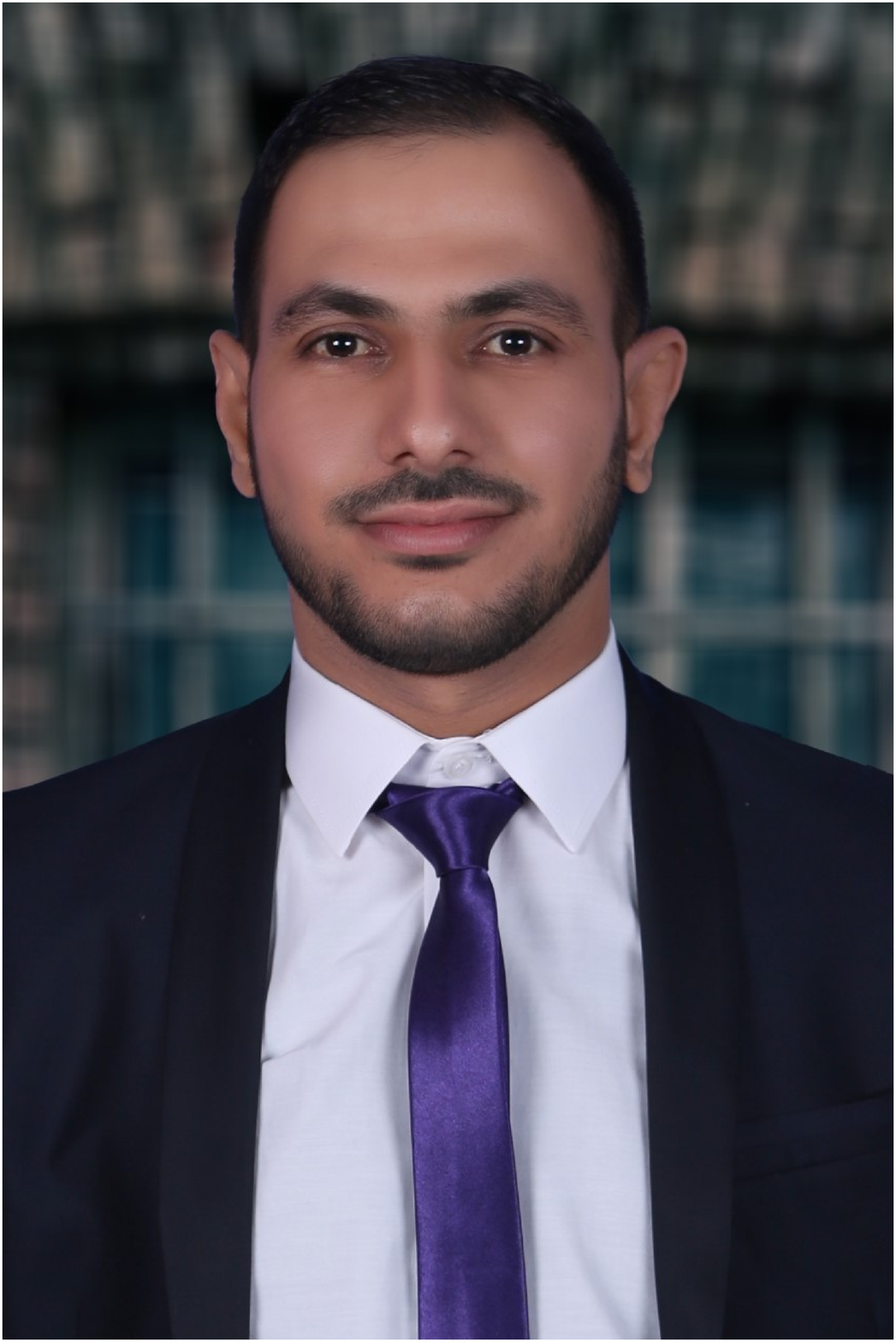}}]{Abdul Karim Gizzini} (Student Member, IEEE)
received the bachelor degree in computer and communication engineering (B.E) from the IUL university of Lebanon, in 2015 followed by the M.E degree in 2017. His master thesis
was hosted by the lebanese national council for scientific research
(CNRS). Since 2019 he is pursuing the Ph.D. degree in wireless communication engineering in ETIS laboratory which is a joint research unit at CNRS (UMR 8051), ENSEA, and CY Cergy Paris University in France. His Ph.D. thesis focuses mainly on deep learning based channel estimation in high mobility vehicular scenarios.
\end{IEEEbiography}

\vskip 0pt plus -1fil


\begin{IEEEbiography}[{\includegraphics[width=1in,height=1.25in,clip,keepaspectratio]{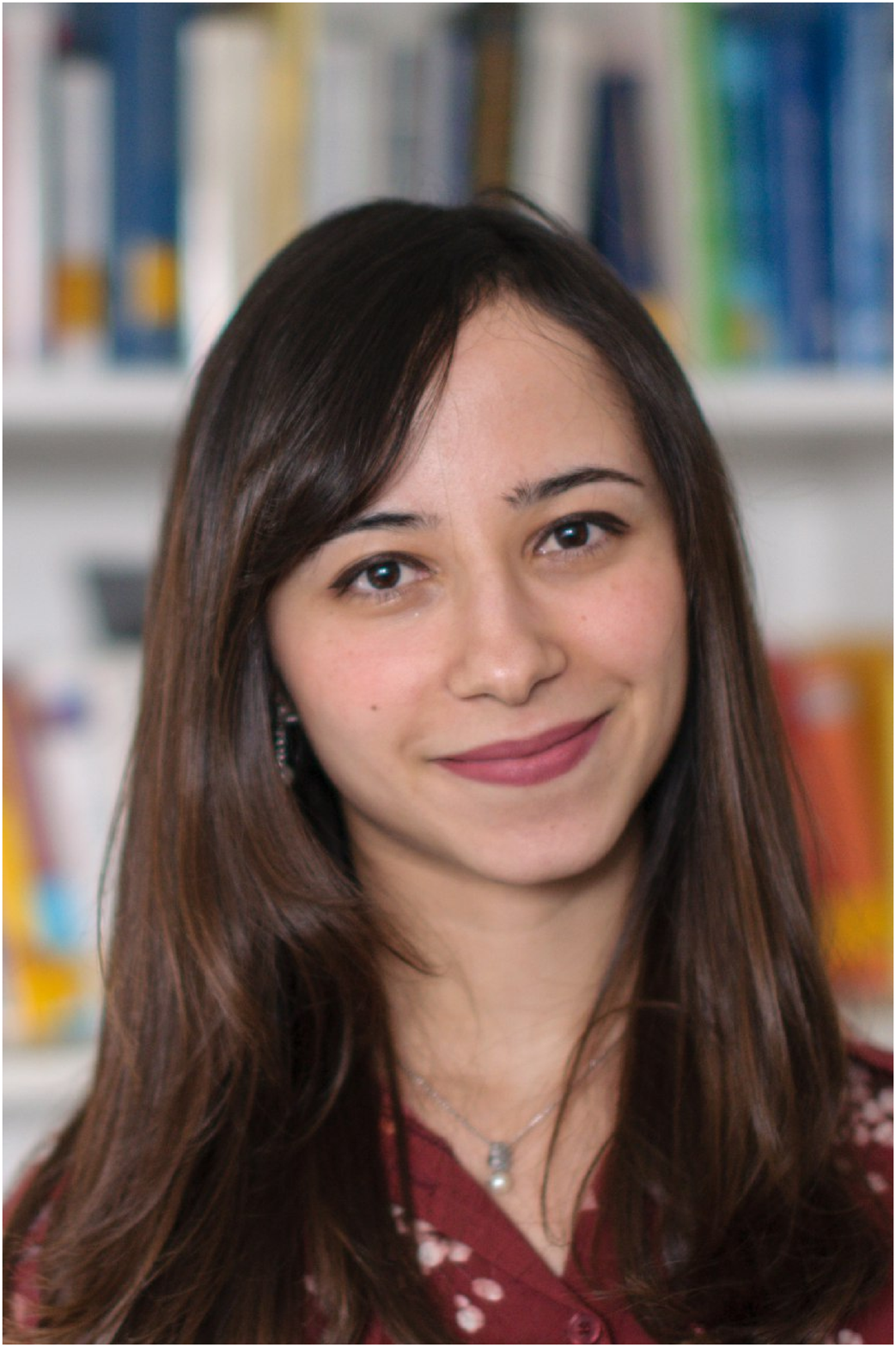}}]{Marwa Chafii}  (Member, IEEE) received her Ph.D. degree in electrical engineering in 2016 from CentraleSupélec, France. She joined the TU Dresden, Germany, in 2018 as a group leader, and ENSEA, France, in 2019 as an associate professor where she held a Chair of Excellence on Artificial Intelligence from CY Initiative. Since September 2021, she is an associate professor at New York University (NYU) Abu Dhabi, and a global network associate professor at NYU WIRELESS, Tandon School of Engineering, NYU. She received the prize of the best Ph.D. in France in the fields of Signal, Image $\&$ Vision, and she has been nominated in the top 10 Rising Stars in Computer Networking and Communications by N2Women in 2020.  She served as Associate Editor at IEEE Communications Letters 2019-2021, where she received the Best Editor Award in 2020. She is currently vice-chair of the IEEE ComSoc ETI on Machine Learning for Communications and leading the Education working group of the ETI on Integrated Sensing and Communications.
\end{IEEEbiography}

\vskip 0pt plus -1fil

\begin{IEEEbiography}[{\includegraphics[width=1in,height=1.25in,clip,keepaspectratio]{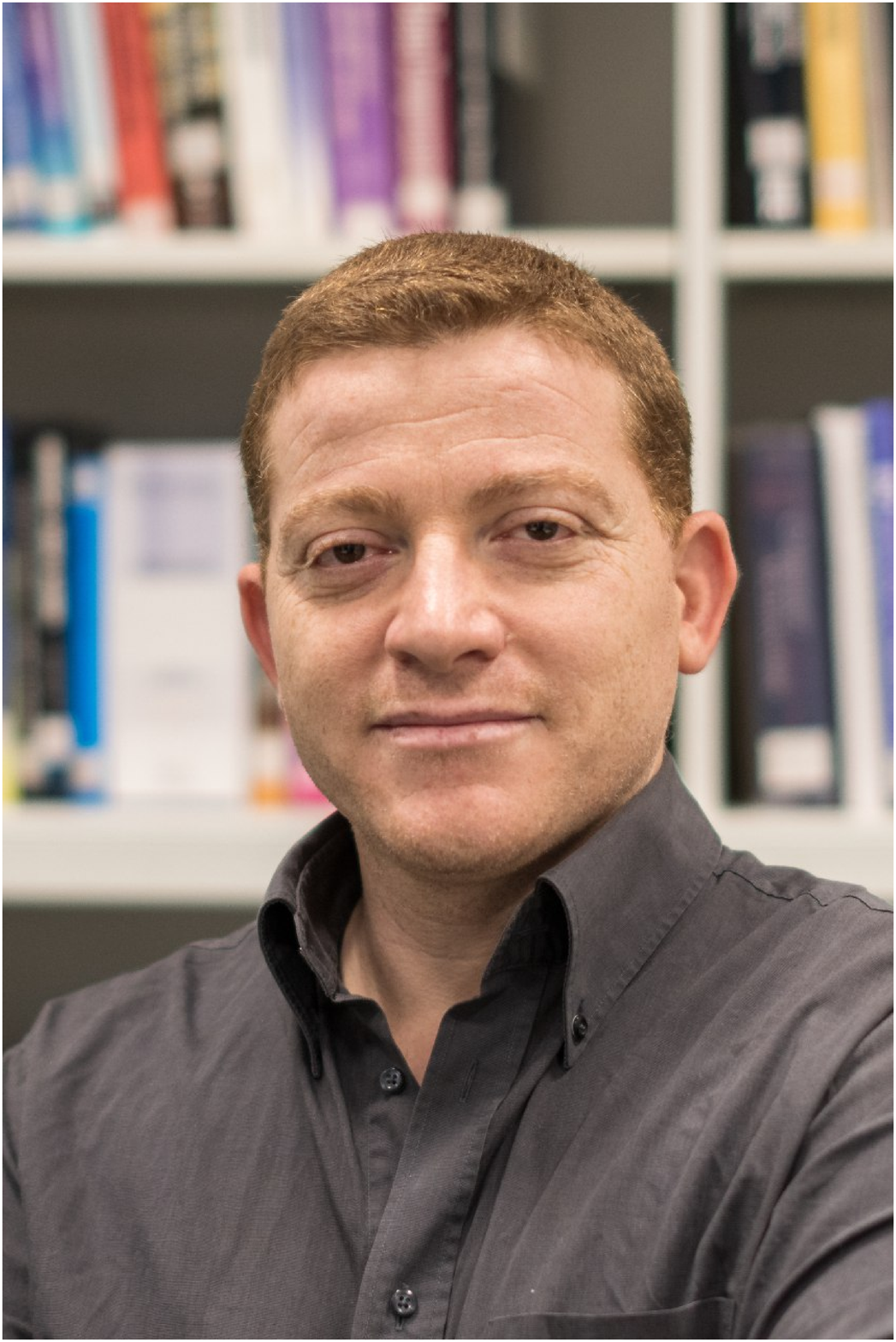}}]{Ahmad Nimr}  (Member, IEEE) received the Ph.D. degree in electrical engineering from TU Dresden, Germany in 2021, M.Sc degree from TU Ilmenau, Germany in in 2014, and the Diploma from HIAST, Syria in 2004. From 2005 to 2011, he pursued industrial carrier in software and hardware development. Since 2020 he has been a research group leader at Vodafone Chair Mobile Communications, TU Dresden. He has worked on several EU and German funded projects with publications in journals and conferences proceedings. His current research focuses on signal processing for communications and sensing from theory to implementation.
\end{IEEEbiography}

\vskip 0pt plus -1fil

\begin{IEEEbiography}[{\includegraphics[width=1in,height=1.25in,clip,keepaspectratio]{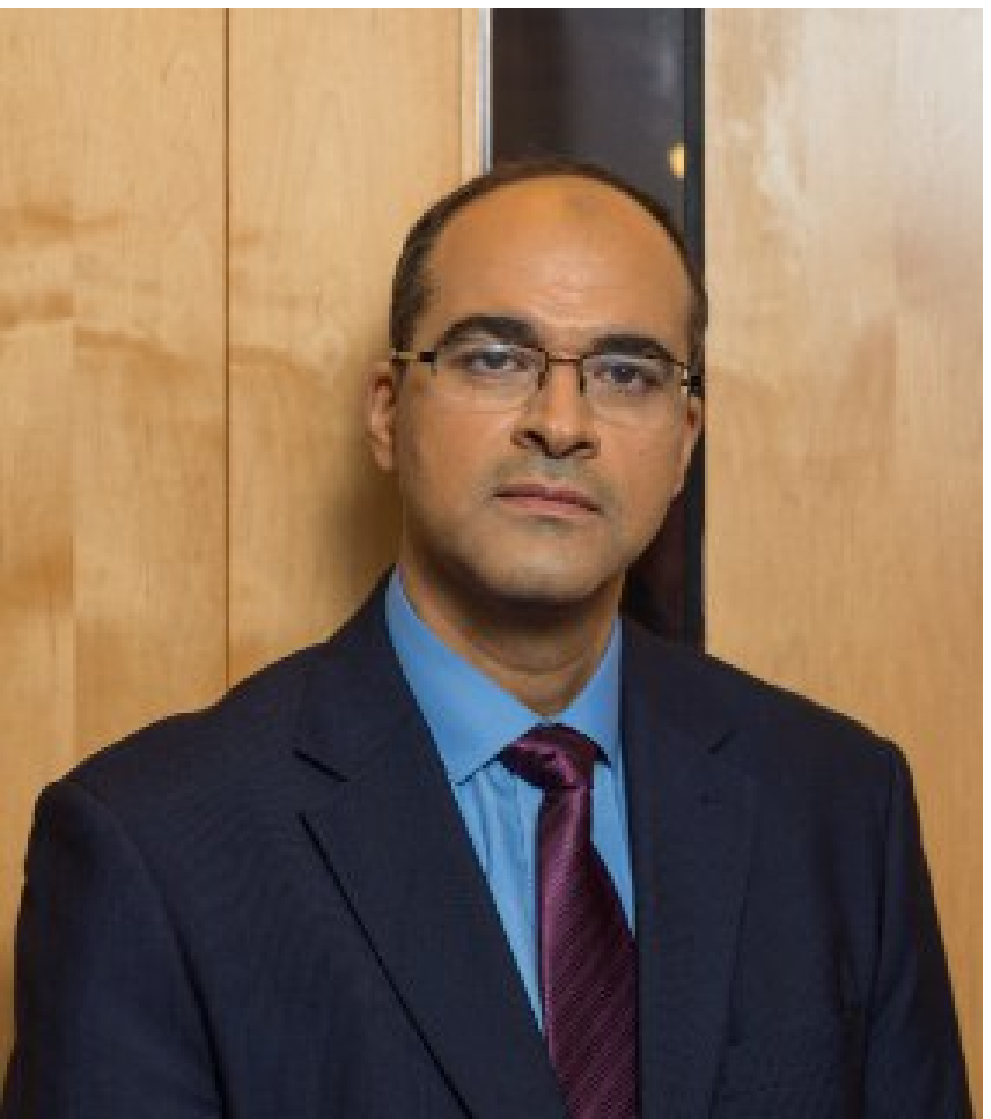}}]{Raed M. Shubair}
(Senior Member, IEEE) received his Ph.D. degree in electrical engineering from the University of Waterloo, Canada, in 1993. He is a Full Professor affiliated to New York University (NYU) Abu Dhabi. He is recipient of the Distinguished Service Award from ACES Society and from MIT Electromagnetics Academy. He is a Board Member of the European School of Antennas, Regional Director for the IEEE Signal Processing Society in Middle East and served as the founding chair of the IEEE Antennas and Propagation Society Educational Initiatives Program. He is a Fellow of MIT Electromagnetics Academy and a Founding Member of MIT Scholars of the Emirates. He is Editor for the IEEE Journal of Electromagnetics, RF, and Microwaves in Medicine and Biology, and Editor for the IEEE Open Journal of Antennas and Propagation. He is a Founding Member of five IEEE society chapters in UAE. He is the Founder and Chair of IEEE at NYUAD. He is an officer for IEEE ComSoc ETI Machine Learning for Communications. He is the founding director of IEEE UAE Distinguished Seminar Series Program.
\end{IEEEbiography}

\vskip 0pt plus -1fil

\begin{IEEEbiography}[{\includegraphics[width=1in,height=1.25in,clip,keepaspectratio]{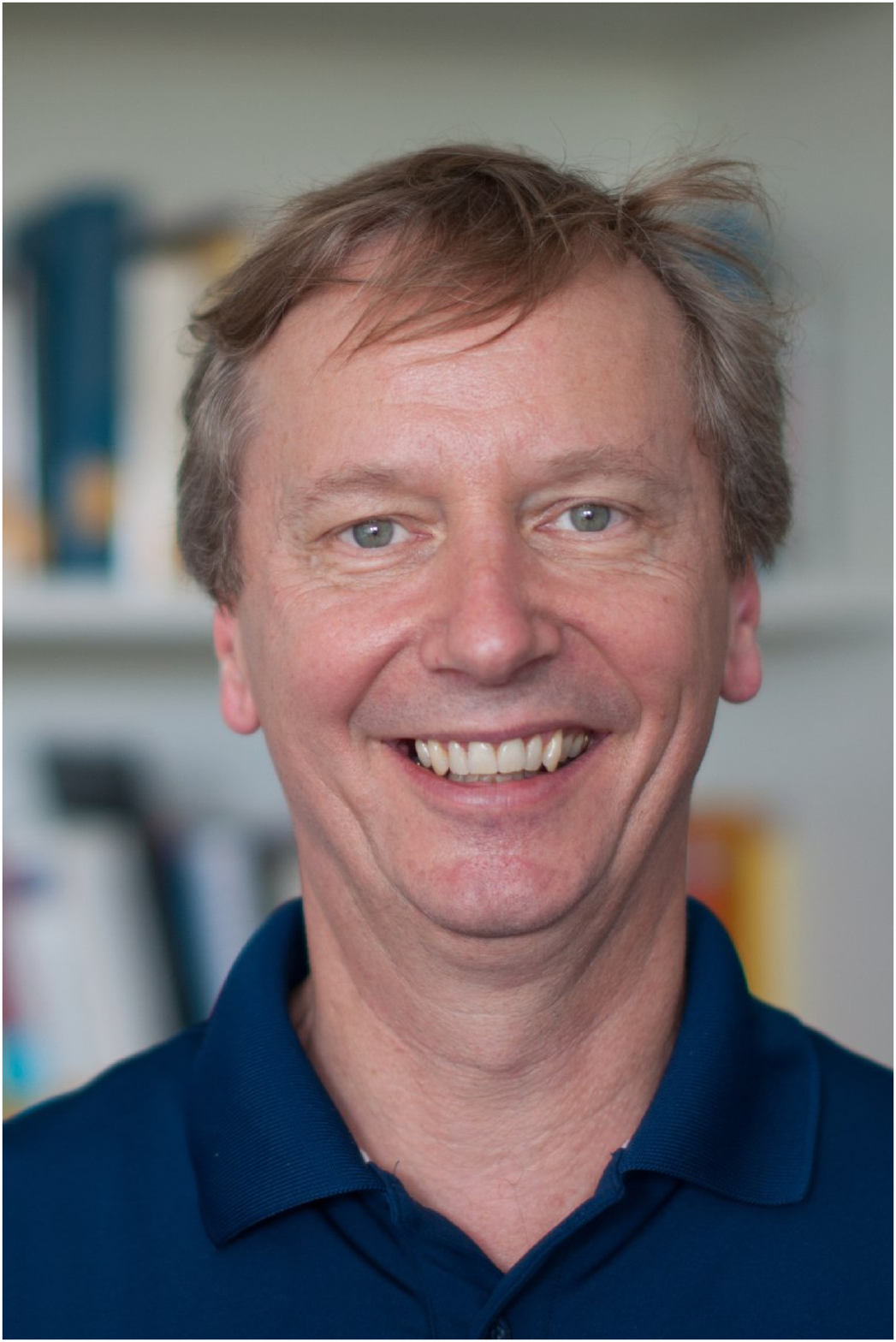}}]{Gerhard Fettweis} (Fellow, IEEE)
is Vodafone Chair Professor at
TU Dresden since 1994, and heads the Barkhausen
Institute since 2018, respectively. He earned his
Ph.D. from RWTH Aachen in 1990. After one year at IBM Research in San Jose, CA, he moved to TCSI Inc., Berkeley,
CA. He coordinates the 5G Lab Germany, and 2
German Science Foundation (DFG) centers at TU
Dresden, namely cfaed and HAEC. His research
focusses on wireless transmission and chip design
for wireless/IoT platforms, with 20 companies from
Asia/Europe/US sponsoring his research. Gerhard is IEEE Fellow, member of the German Academy of Sciences (Leopoldina), the German Academy of
Engineering (acatech), and received multiple IEEE recognitions as well has the VDE ring of honor. In Dresden his team has spun-out sixteen start-ups,
and setup funded projects in volume of close to EUR 1/2 billion. He co-chairs the IEEE 5G Initiative, and has helped organizing IEEE conferences, most
notably as TPC Chair of ICC 2009 and of TTM 2012, and as General Chair of VTC Spring 2013 and DATE 2014.
\end{IEEEbiography}

\end{document}

%% file: commands.tex
 \usepackage{bm}

\newcommand{\figref}[1]{Fig.~\ref{#1}}

\newcommand{\tabref}[1]{Table~\ref{#1}}

\newcommand{\LeftP} {\left\lbrace }
\newcommand{\RightP}{\right\rbrace }

\newcommand{\compl}{\mathbb{C}}         

\newcommand{\ma}  [1]{ \bm{#1} } 

\newcommand{\Ex}[1]{\mathrm{E}\left[ #1\right]} 
\newcommand{\trace}[1] {\mathrm{trace}\LeftP #1 \RightP }

\newcommand{\Vect}  [1] {\mathrm{vec}  \LeftP #1 \RightP } 

\newcommand{\set} [1]{{\mathcal {#1}}} 
\newcommand{\Kon} {\set{K}_{\text{on}}} 
\newcommand{\Kp} {\set{K}_{\text{p}}} 

\newcommand{\Kf} {\set{K}_{f}} 
\newcommand{\Kt} {\set{K}_{t}} 



%% file: acronyms.tex
\begin{acronym}
    \acro{C-ITS}{cooperative intelligent transportation system}
    \acro{RBF}{radial basis function}
    \acro{CNN}{Convolutional neural network}
    \acro{TS-ChannelNet}{Temporal spectral ChannelNet}
    \acro{TDR}{transmission data rate}
    \acro{LSTM}{long short-term memory}
    \acro{ALS}{accurate LS}
    \acro{SLS}{simple LS}
    \acro{ChannelNet}{channel network}
    \acro{ADD-TT}{average decision-directed with time truncation}
    \acro{WI}{weighted interpolation}
    \acro{DD}{decision-directed}
    \acro{SR-ConvLSTM}{super resolution convolutional long short-term memory}
    \acro{SR-CNN}{super resolution CNN}
    \acro{DN-CNN}{denoising CNN}
    \acro{V2V}{vehicle-to-vehicle}
    \acro{V2I}{vehicle-to-infrastructure}
    \acro{VTV}{vehicle-to-vehicle}
    \acro{RTV}{roadside-to-vehicle}
    \acro{DPA}{data-pilot aided}
    \acro{STA}{spectral temporal averaging}
    \acro{M-STA}{modified STA}
    \acro{LMMSE}{linear minimum mean-squared error}
    \acro{M-CDP}{modified CDP}
    \acro{M-TRFI}{modified TRFI}
    \acro{SNR}{signal-to-noise ratio}
    \acro{CDP}{Constructed data pilots}
    \acro{DFT}{discrete Fourier transform}
    \acro{T-DFT}{truncated discrete Fourier transform}
    \acro{TA-MDFT}{temporal averaging M-DFT}
    \acro{TA-DFT}{TA-DFT}
    \acro{STS}{short training symbols}
    \acro{LTS}{long training symbols}
    \acro{SS}{signal symbol}
    \acro{TDL}{tapped delay line}
    \acro{TRFI}{time domain reliable test frequency domain interpolation}
    \acro{SBS}{symbol-by-symbol}
    \acro{FBF}{frame-by-frame}
    \acro{MMSE-VP}{minimum mean square error using virtual pilots}
    \acro{DL}{deep learning}
    \acro{AE-DNN}{auto-encoder deep neural network}
    \acro{DNN}{deep neural networks} 
    \acro{ISI}{inter-symbol-interference}
    \acro{OFDM}{orthogonal frequency-division multiplexing}
    \acro{LS}{least square}
    \acro{RS}{reliable subcarriers}
    \acro{URS}{unreliable subcarriers}
    \acro{MMSE}{minimum mean squared error}
    \acro{SoA}{state of the art}
    \acro{NMSE}{normalized mean-squared error}
	\acro{BER}{bit error rate}
	\acro{RSU}{road side unit}
	\acro{AE}{auto-encoder}
    \acro{CP}{cyclic prefix}
\acro{PLCP}{physical layer convergence protocol}
\acro{MSE}{mean squared error}
\end{acronym}

%% file: abstract.tex
\begin{abstract}
IEEE 802.11p standard defines wireless technology protocols that enable vehicular transportation and manage traffic efficiency. A major challenge in the development of this technology is ensuring communication reliability in highly dynamic vehicular environments, where the wireless communication channels are doubly selective, thus making channel estimation and tracking a relevant problem to investigate. In this paper, a novel deep learning (DL)-based weighted interpolation estimator is proposed to accurately estimate vehicular channels especially in high mobility scenarios. The proposed estimator is based on modifying the pilot allocation of the IEEE 802.11p standard so that more transmission data rates are achieved. Extensive numerical experiments demonstrate that the developed estimator significantly outperforms the recently proposed DL-based frame-by-frame estimators in different vehicular scenarios, while substantially reducing the overall computational complexity.
\end{abstract}


%% file: introduction.tex
\section{Introduction} \label{introduction}
In practical vehicular applications, the communication reliability between vehicles is very important. In general, communication reliability can be evaluated by the accurate data recovery of the transmitted data at the receiver which depends primarily on the channel estimation accuracy. A precisely-estimated channel is critical for the equalization, demodulation, and decoding operations to follow. Hence, an accurate and robust channel estimation is essential for improving the overall system performance. Accurate channel estimation in vehicular communication is a challenging task, since in vehicular environments wireless channels have a high Doppler shift and a large delay spread~\cite{ref_V2X_Channels,bomfin2021robust}. As a result, the channel shows time and frequency selective fading characteristics due to motion and multi-path components. This double selectivity arises in high mobility vehicular environment where the channel rapidly varies within the transmitted frame.

IEEE 802.11p frame structure~\cite{ref_IEEE_Spec} allocates two full preamble symbols that are used for basic {\ac{LS}} estimation, and four pilot subcarriers within the transmitted data symbols which are used for channel variation tracking over time. This basic estimation at the beginning of the frame is simple, but it becomes invalid for the equalization of the successive transmitted symbols as a result of high channel variation in vehicular environments. To improve the basic {\ac{LS}} estimation performance, there exist two channel estimation categories: (\textit{i}) \ac{SBS} estimators, where the channel is estimated for each received symbol separately~\cite{ref_STA,ref_CDP,ref_TRFI}. (\text{ii}) \ac{FBF} estimators, where the previous, current and future pilots are employed in the channel estimation for each received symbol \cite{ehsanfar2020uw}. The well known {\ac{FBF}} estimator is the conventional 2D \ac{LMMSE} where the channel and noise statistics are utilized in the estimation, and thus, leading to comparable performance to the ideal case. However, the 2D {\ac{LMMSE}} suffers from high computational complexity making it impractical in real case scenarios. Therefore, there is a  need for robust and low complexity {\ac{FBF}} estimators.

Recently, {\ac{DL}} algorithms have been integrated into wireless communications physical layer applications~\cite{ref_DL_PHY1, ref_DL_PHY2} including channel estimation~\cite{ref_DL_Chest1,ref_DL_Chest2,ref_DL_Chest3}, due to its great success in improving the overall system performance, especially when used on top of low-complexity conventional estimators. {\ac{DL}} algorithms are characterized by robustness, low-complexity, and good generalization ability making the integration of {\ac{DL}} into communication systems beneficial. Motivated by these advantages, {\ac{DL}} algorithms have been integrated in IEEE 802.11p channel estimators in two different manners: (\textit{i}) Feed-forward deep neural networks with different architectures and configurations are employed on top of {\ac{SBS}} estimators~\cite{ref_AE_DNN,ref_STA_DNN,ref_TRFI_DNN}, where the channel is estimated for each received data symbol directly. (\textit{ii}) {\acp{CNN}} are integrated in the {\ac{FBF}} estimators as a good alternative to the 2D {\ac{LMMSE}} estimator, where the estimated channel for the whole frame is considered as a 2D low resolution noisy image and \ac{CNN}-based processing is applied as super resolution and denoising techniques. Unlike the 2D {\ac{LMMSE}} estimator, the {\ac{CNN}} processing achieves considerable performance gain while preserving low computational complexity. The higher performance accuracy can be achieved by utilizing {\ac{FBF}} estimators, since the channel estimation of each symbol takes advantage from the knowledge of previous, current, and future allocated pilots within the frame. Unlike, {\ac{SBS}} estimators, where only the previous and current pilots are exploited in the channel estimation for each received symbol. In this paper, we will focus on {\ac{FBF}} estimators, especially, the recently proposed {\ac{CNN}}-based ones. 

In~\cite{ref_ChannelNet}, the authors propose {\ac{CNN}}-based channel estimator denoted as {\ac{ChannelNet}}, that applies {\ac{RBF}} interpolation as an initial channel estimation, after that the {\ac{RBF}} estimated channel is considered as a low resolution image, where {\ac{SR-CNN}} followed by {\ac{DN-CNN}} are integrated on top of the {\ac{RBF}} estimated channel. {\ac{ChannelNet}} estimator suffers mainly from high computational complexity and considerable performance degradation in high mobility vehicular scenarios. {\ac{TS-ChannelNet}} has been proposed in~\cite{ref_TS_ChannelNet}, where {\ac{ADD-TT}} interpolation that is based on the demodulation and averaging of each received symbol. After that, {\ac{SR-ConvLSTM}} is used to improve {\ac{ADD-TT}} interpolation accuracy. {\ac{TS-ChannelNet}} estimator outperforms the {\ac{ChannelNet}} estimator especially in low {\ac{SNR}} regions. Moreover, the {\ac{TS-ChannelNet}} estimator has lower computational complexity than the {\ac{ChannelNet}} estimator since only one {\ac{CNN}} network is considered instead of two {\acp{CNN}} as proposed in the {\ac{ChannelNet}} estimator. Nevertheless, {\ac{TS-ChannelNet}} also suffers from a considerable performance degradation in high mobility vehicular scenarios. It is worth mentioning that both {\ac{ChannelNet}} and {\ac{TS-ChannelNet}} estimators suffer from high latency at the receiver, since the receiver should wait for the arrival of the whole frame in order to start the channel estimation process. 

In order to overcome the high complexity and performance degradation in high vehicular scenarios, we propose in this paper novel hybrid and adaptive {\ac{WI}} channel estimators that utilize new pilot allocation schemes for the IEEE 802.11p that insert pilot symbols into the transmitted frame. Unlike {\ac{ChannelNet}} and {\ac{TS-ChannelNet}} estimators, where the applied interpolation techniques are performed in time and frequency, the developed {\ac{WI}} estimators require only weighted time interpolation of the inserted pilot symbols. The number of inserted pilots is controlled and adapted according to the mobility condition. In particular, one \ac{OFDM} symbol with all pilots is inserted at the end of the transmitted frame in low mobility scenario and as the mobility increases, more  pilot \ac{OFDM} symbols are required. All the other symbols within the transmitted frame are fully allocated by data subcarriers and considered as data symbols. The estimated channel for the data symbols is considered as a weighted summation of the estimated channels at the inserted pilot symbols. According to the used frame structure, the receiver treats the received frame as subframes, thus achieving lower latency at the receiver, since the channel estimation starts upon receiving each subframe instead of waiting for the whole frame. Moreover, in order to gain more {\ac{TDR}}, few pilots can be inserted within the pilot \ac{OFDM} symbols depending on the channel delay profile.

Further performance improvement can be achieved by integrating optimized {\ac{SR-CNN}} or {\ac{DN-CNN}} as post processing modules after the {\ac{WI}} estimators. The proposed {\ac{WI}} estimators enjoy low-complexity and robustness, and achieve good performance in high-mobility vehicular scenarios. Additionally, the proposed {\ac{WI}} estimators contribute in latency reduction at the receiver, besides gaining more transmission data rates. 

To the best of our knowledge, there is no recent {\ac{FBF}} estimators that modify the IEEE 802.11p pilot allocation, however, the work proposed in~\cite{mid1,mid2} employ modified frame structure but for {\ac{SBS}} channel estimation, where a midamble pilot symbol is inserted frequently within the transmitted frame and used for estimating the channel for all successive symbols.

The contributions of this paper can be summarized as follows:
\begin{itemize}
    \item Proposing a hybrid, adaptive, low-complexity, and robust {\ac{WI}} channel estimators that modify the pilot allocation schemes of the IEEE 802.11p within the transmitted frame and adapt the employed scheme according to the mobility condition.
    
    \item Deriving analytically the expression of the employed interpolation matrix for the proposed pilot allocation schemes.
    
    \item Providing a brief overview of the {\ac{CNN}} networks, especially those used in the state-of-the-art of the channel estimation, such as {\ac{SR-CNN}}, {\ac{DN-CNN}}.
    
    \item Integrating an optimized {\ac{SR-CNN}} and {\ac{DN-CNN}} networks on top of the {\ac{WI}} estimators to enhance the {\ac{BER}} and {\ac{NMSE}} performance.
    
    \item Showing that the proposed {\ac{WI}} estimators outperform {\ac{ChannelNet}} and {\ac{TS-ChannelNet}} estimators in terms of latency and transmission data rates.
    
    \item Providing a detailed computational complexity analysis for the studied channel estimators, where we show that the proposed {\ac{WI}} estimators outperform the benchmarked estimators with substantial reduction in complexity.
\end{itemize}

The remainder of this paper is organized as follows: in Section~\ref{system_model}, the IEEE 802.11p standard and the system model are described. Section~\ref{literature_review} provides a detailed description of the recently proposed {\ac{ChannelNet}} and {\ac{TS-ChannelNet}} estimators. The proposed {\ac{WI}} estimators, as well as the analytical interpolation matrices derivations are presented in Section~\ref{WI}. Then, a detailed overview of the {\ac{CNN}} based channel estimation and the proposed optimized architectures is provided in Section~\ref{CNN}. In Section~\ref{results}, simulation results are presented for different vehicular channel models conditions using different modulation orders, where the performance of the proposed estimators is evaluated in terms of \ac{BER}, \ac{NMSE}, {\ac{TDR}}, and latency. Detailed computational complexity analysis is provided in Section~\ref{complexity}. Finally, conclusions are given in Section~\ref{conclusions}. 

%% file: system_model.tex
\section{System Model Description} \label{system_model}

IEEE 802.11p is an international standard that manages wireless access in vehicular environments. IEEE 802.11p defines the communications between high-speed vehicles ({\ac{V2V}}) and between the vehicles and the roadside infrastructure ({\ac{V2I}}) in the licensed intelligent transportation systems band~\cite{V2x_Comm}. IEEE 802.11p uses {\ac{OFDM}} transmission scheme with $K = 64$ total subcarriers. $K_{\text{on}} = 52$ active subcarriers are used, and they are divided into $K_{\text{d}} = 48$ data subcarriers and $K_{\text{p}} = 4$ pilot subcarriers. The remaining $12$ subcarriers are used as a guard band. IEEE 802.11p standard employs two \ac{LTS} preamble symbols at the beginning of the transmitted frame that are used for signal detection and channel estimation at the receiver. Moreover, it supports transmitting data at different data rates employing different modulation orders. Table~\ref{tb:IEEE80211p_specs} shows the IEEE 802.11p physical layer main specifications. A detailed discussion of the IEEE 802.p standard and all its features is presented in~\cite{ref_IEEE_Spec}.

In this paper, we assume perfect synchronization at the receiver, and we consider a frame that consists of two {\ac{LTS}} at the beginning followed by $I$ OFDM data symbols. The received {\ac{OFDM}} symbol $\tilde{\ma{y}}_i[k]$ can be expressed in terms of the transmitted {\ac{OFDM}} symbol $\tilde{\ma{x}}_i[k]$ as follows
\begin{equation}
\tilde{\ma{y}}_i[k] = \tilde{\ma{h}}_i[k] \tilde{\ma{x}}_i[k] + \tilde{\ma{v}}_i[k],~ k \in \Kon.
\label{eq:datasymbols}
\end{equation}

Here, $\tilde{\ma{v}}_i[k]$ denotes the noise of the $k$-th subcarrier in the $i$-th \ac{OFDM} symbol. $\tilde{\ma{h}}_i[k]$ represents the time variant frequency response of the channel for the $i$-th \ac{OFDM} symbol. Moreover,  let $\tilde{\ma{Y}}[k,i] \in \compl ^{K_{\text{on}}\times I}$ and $\tilde{\ma{X}}[k,i] \in \compl ^{K_{\text{on}} \times I}$ be the received and transmitted {\ac{OFDM}} frame respectively. Then,~\eqref{eq:datasymbols} can be expressed in a matrix form as follows

\begin{equation}
\tilde{\ma{Y}}[k,i] = \tilde{\ma{H}}[k,i]  \tilde{\ma{X}}[k,i] + \tilde{\ma{V}}[k,i],~ k \in \Kon,
\label{eq:preamble_freq}
\end{equation}
where $\tilde{\ma{V}}[k,i] \in \compl ^{K_{\text{on}}\times I}$ and $\tilde{\ma{H}} \in \compl ^{K_{\text{on}}\times I}$ denote the noise and the time variant frequency response of the channel for all symbols within the transmitted \ac{OFDM} frame respectively.

\begin{table}[t]
	\renewcommand{\arraystretch}{1.3}
	\small
	\centering
	\caption{IEEE 802.11p physical layer  specifications.}
	\label{tb:IEEE80211p_specs}
	\begin{tabular}{|c|c|}
		\hline
		\textbf{Parameter}      & \textbf{IEEE 802.11p}                    \\ \hline
		Bandwidth               & 10 MHz  \\ \hline
		Guard interval duration & 1.6~$\mu\mbox{s}$       \\ \hline
		Symbol duration         & 8~$\mu\mbox{s}$                 \\ \hline
		Short training symbol duration         & 1.6~$\mu\mbox{s}$    \\ \hline
		Long training symbol duration         & 6.4~$\mu\mbox{s}$ \\ \hline
		Total subcarriers      & 64     \\ \hline
		Pilot subcarriers      & 4  \\ \hline
		Data subcarriers       & 48     \\ \hline
		Null subcarriers       & 12      \\ \hline
		Subcarrier spacing     & 156.25 KHz       \\ \hline
	\end{tabular}
\end{table}

%% file: SoA_estimators.tex
\section{SoA of FBF channel estimators}\label{literature_review}
In the section, the conventional 2D {\ac{LMMSE}} estimator and  recently proposed {\ac{CNN}}-based estimators schemes are presented and discussed.

\subsection{Conventional 2D LMMSE estimator}

The basic idea of the conventional 2D {\ac{LMMSE}} estimation is to employ the channel correlation matrices and the noise power in the channel estimation of each data subcarrier within the received {\ac{OFDM}} frame, such that
\begin{equation}
\hat{\tilde{\ma{h}}}_{{\text{LMMSE}}_{i}} =  \ma{R}_{{\tilde{\ma{h}}}_{i}, \tilde{\ma{h}}_{p}} \big( \ma{R}_{\tilde{\ma{h}}_{p}, \tilde{\ma{h}}_{p}} + \sigma^{2} \ma{I}^{\prime} \big)^{-1} \bar{{\ma{h}}}_{\text{LS}}.
\label{eq:lmmse1}
\end{equation}
$\ma{R}_{{\tilde{\ma{h}}}_{i}, \tilde{\ma{h}}_{p}} \in \compl ^{ K_{p} I \times 1}$ represents the $k$-th column of the cross correlation matrix between the real channel vectors $\tilde{\ma{h}}_{i}$ and $\tilde{\ma{h}}_{p}$ at the data and the pilot subcarriers within the received {\ac{OFDM}} frame respectively. Moreover, $\ma{R}_{\tilde{\ma{h}}_{p}, \tilde{\ma{h}}_{p}}$ denotes the auto correlation matrix of $\tilde{\ma{h}}_{p}$, $I^{\prime} \in \mathbb{R}^{K_{p} I \times K_{p} I} $ is the identity matrix, and $\bar{{\ma{h}}}_{\text{LS}} = \Vect{\hat{\tilde{\ma{H}}}_{\text{LS}}} \in \compl ^{K_{p}I \times 1}$ represents the {\ac{LS}} estimated channel vector at the pilot subcarriers within the received {\ac{OFDM}} frame, where
\begin{equation}
\hat{\tilde{\ma{H}}}_{\text{LS}}[k,i] = \frac{\tilde{\ma{Y}}[k,i]}{\tilde{\ma{P}}[k,i]},~ k \in {\set{K}}_{\text{p}},~ 1\leq i \leq I.
\label{eq: LS-RBF}
\end{equation}
    
$\tilde{\ma{P}}[k,i]$ is the frequency domain pre-defined pilot subcarriers and ${\set{K}}_{\text{p}}$ denotes the allocated sparse pilots indices within the received {\ac{OFDM}} symbol.

The conventional 2D {\ac{LMMSE}} estimator achieves almost similar performance as the ideal channel, however it suffers from high computational complexity, therefore,  making the it impractical in real-time scenario. In contrast, CNN provides a good performance-complexity trade-off by learning the channel statistics while recording an acceptable low computational complexity. Therefore, it is employed in the {\ac{FBF}} channel estimation as an alternative to the 2D {\ac{LMMSE}} estimator.

\subsection{ChannelNet estimator}

In~\cite{ref_ChannelNet}, the authors propose a CNN-based channel estimator denoted as {\ac{ChannelNet}} scheme, where 2D \ac{RBF} interpolation is applied as an initial channel estimation. The basic motivation of the 2D {\ac{RBF}} interpolation is to approximate multidimensional scattered unknown data, from their surrounding neighbors known data, employing the radial basis function~\cite{ref_RBF}. To do so, the distance function is calculated between every data point to be interpolated and its neighbours, where closer neighbors are assigned higher weights. After that, the \ac{RBF} interpolated frame is considered as a low resolution image, where {\ac{SR-CNN}} is utilized to get better estimation. Finally, in order to alleviate the impact of noise within the high resolution estimated frame, \ac{DN-CNN} is implemented on top of the \ac{SR-CNN} resulting in a high resolution, noise alleviated estimated channels. The ChannelNet estimator considers sparsed allocated pilots within the IEEE 802.11p frame and it first applies the {\ac{LS}} estimation to the pilot subcarriers within the received {\ac{OFDM}} frame as illustrated in~{\eqref{eq: LS-RBF}}. After that, The 2D {\ac{RBF}} interpolation is obtained by the weighted summation of the distance between each data subcarrier to be interpolated and all the pilot subcarriers within the received {\ac{OFDM}} frame, such that
\begin{equation}
\hat{\tilde{\ma{H}}}_{\text{RBF}}[k,i] = \sum_{j = 1}^{{K_{{p}}}I} \omega_j \Phi (| k - \Kf[j]| ,  |i - \Kt[j]|).
\label{eq:RBF_eq}    
\end{equation}

\begin{figure*}[t]
	\centering
	\includegraphics[height=8cm,width=2\columnwidth]{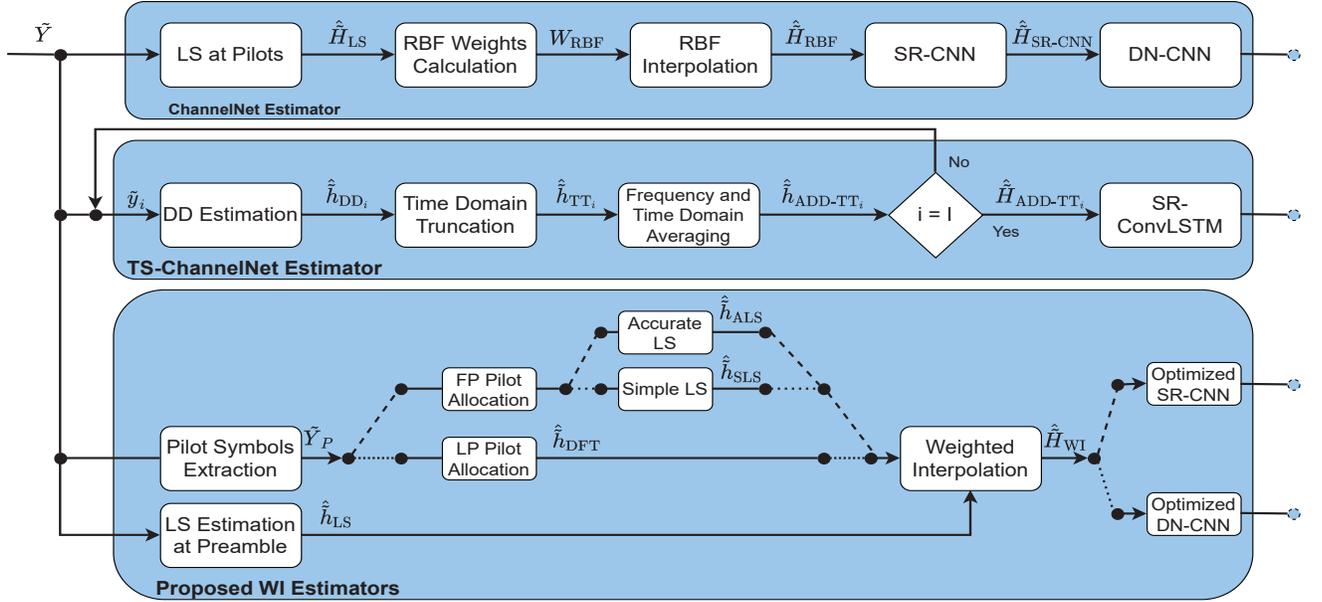}
	\caption{Proposed {\ac{WI}} estimators block diagram.}
	\label{fig:soa_Figure}
\end{figure*}

$\Kf = [{\set{K}}_{\text{p}_{1}}, \dots, {\set{K}}_{\text{p}_{I}}] \in \mathbb{R}^{1 \times K_{p}I} $ and $\Kt = [(1)_{\times K_{p}},\dots, (I)_{\times K_{p}}]  \in \mathbb{R}^{1 \times K_{p}I}$ denote the frequency and time indices vectors of the sparsed allocated pilot subcarriers within the received {\ac{OFDM}} frame. $\omega_j$ represents the {\ac{RBF}} weight multiplied by the {\ac{RBF}} interpolation function $\Phi(.)$ between the $(k, i)$ data subcarrier and the $(\Kf[j], \Kt[j])$ pilot subcarrier. In~\cite{ref_ChannelNet}, the {\ac{RBF}} gaussian function is applied, such that
\begin{equation}
    \Phi(x,y) = e^{-\frac{(x + y)^2}{r_{0}}}, \label{eq:RBF_PHI}
\end{equation}
where $r_{0}$ is the 2D {\ac{RBF}} scale factor and it varies according to the used {\ac{RBF}} function. We note that changing the value of $r_{0}$ changes the shape of the interpolation function. Moreover, the \ac{RBF} weights $\ma{w}_{\text{RBF}} = [\omega_{1}, \dots, \omega_{K_{p}I}] \in \mathbb{R}^{K_{p}I \times 1}$ are calculated using the following relation
\begin{equation}
    \ma{A}_{\text{RBF}} \ma{w}_{\text{RBF}} = \bar{{\ma{h}}}_{\text{LS}},
\label{eq:RBF_w}
\end{equation}
where $\ma{A}_{\text{RBF}} \in \mathbb{R}^{ {K_{{p}}}I  \times {K_{{p}}}I }$ is the {\ac{RBF}} interpolation matrix  of the pilots subcarriers, with entries $a_{i,j} = \Phi(\Kf[i],\Kt[j])$ where $i,j = 1, \dots, K_{p}I$. After computing $\ma{W}_{\text{RBF}}$, the {\ac{RBF}} estimated channel for every data subcarriers within the received {\ac{OFDM}} frame can be calculated as shown in~\eqref{eq:RBF_eq}. Finally, the {\ac{RBF}} interpolation estimated frame $\hat{\tilde{\ma{H}}}_{\text{RBF}}$ is fed as an input to {\ac{SR-CNN}} and {\ac{DN-CNN}} in order to improve the channel estimation accuracy, and alleviate the noise impact.

The ChannelNet estimator limitations lie in: (\textit{i}) 2D {\ac{RBF}} interpolation high computational complexity that results from the computation of~{\eqref{eq:RBF_w}} for the channel estimation of each data subcarrier. (\textit{ii}) The 2D {\ac{RBF}} function and scale factor should be optimized according to the channel variations. (\textit{iii}) The integrated {\ac{SR-CNN}} and {\ac{DN-CNN}} architectures have considerable computational complexity. We note that, the ChannelNet estimator uses a fixed  {\ac{RBF}} function and scale factor, therefore, it suffers from considerable performance degradation especially in low {\ac{SNR}} regions where the impact of noise is dominant, and high mobility vehicular scenarios, where the channel varies rapidly within the {\ac{OFDM}} frame.
\subsection{TS-\ac{ChannelNet} estimator}
\ac{TS-ChannelNet}~\cite{ref_TS_ChannelNet} 
is based on applying {\ac{ADD-TT}} interpolation to the received {\ac{OFDM}} frame. After that, an accurate estimation is obtained by implementing {\ac{SR-ConvLSTM}} network in order to track vehicular channel variations by learning the time and frequency correlations of the vehicular channel. 
First, the {\ac{LS}} estimation is applied using the two \ac{LTS} received preambles denoted as $\tilde{\ma{y}}_{{\text{LTS}}_{1}}[k]$, and $\tilde{\ma{y}}_{{\text{LTS}}_{2}}[k]$, and the predefined frequency domain preamble sequence $\tilde{\ma{p}}[k]$ such that
\begin{equation}
\hat{\tilde{\ma{h}}}_{\text{LTS}}[k] = \frac{\tilde{\ma{y}}_{{\text{LTS}}_{1}}[k] + \tilde{\ma{y}}_{{\text{LTS}}_{2}}[k]}{2\tilde{\ma{p}}[k]}.
\label{eq: LS1}
\end{equation}
After that, \ac{DD} channel estimation is applied by employing the demapped data subcarriers of the previous received {\ac{OFDM}} symbol to estimate the channel for the current {\ac{OFDM}} symbol. The {\ac{DD}} estimation consists of the following steps

\begin{enumerate}
    \item Equalization: the $i$-th received {\ac{OFDM}} symbol is equalized by the previously {\ac{DD}} estimated channel, such that
    
    \begin{equation}
    {\tilde{\ma{y}}_{\text{eq}_{i}}[k]} = \frac{\tilde{\ma{y}}_i[k]}{\hat{\tilde{\ma{h}}}_{\text{ADD-TT}_{i-1}}[k]}
    ,~ \hat{\tilde{\ma{h}}}_{\text{ADD-TT}_{0}}[k] = \hat{\tilde{\ma{h}}}_{\text{LS}}[k]. \label{eq: ADD_TT_1}
    \end{equation}
    
    \item Demapping: ${\tilde{\ma{y}}_{\text{eq}_{i}}[k]}$ is demapped to the nearest constellation point to obtain $\tilde{\ma{d}}_i[k]$.
    
    \item {\ac{DD}} estimation: update the {\ac{DD}} estimated channel for the $i$-th received {\ac{OFDM}} symbol, by dividing $\tilde{\ma{y}}_i[k]$ with $\tilde{\ma{d}}_i[k]$ as expressed below
    \begin{equation}
    \hat{\tilde{\ma{h}}}_{\text{DD}_{i}}[k]= \frac{\tilde{\ma{y}}_i[k]}{\tilde{\ma{d}}_i[k]}.
    \label{eq: ADD_TT_2}
    \end{equation}
\end{enumerate}

\begin{figure*}[t]
\centering
  \begin{subfigure}[t]{0.25\textwidth} \centering
    \includegraphics[height=0.8\linewidth]{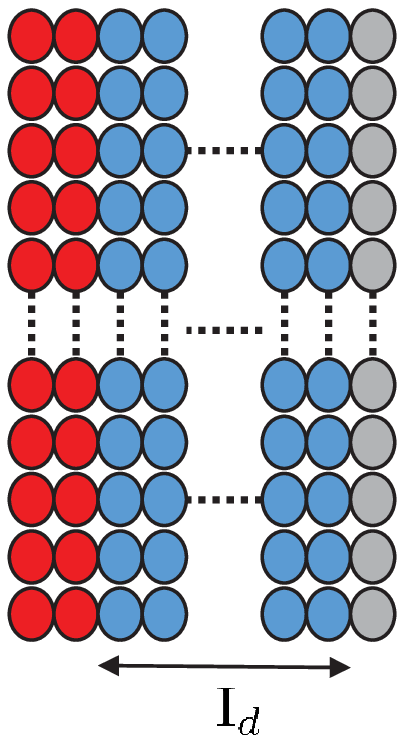}
    \caption{Low mobility.}
    \label{fig:Full-LM}
  \end{subfigure}\hfill
  \begin{subfigure}[t]{0.25\textwidth} \centering
    \includegraphics[height=0.8\linewidth]{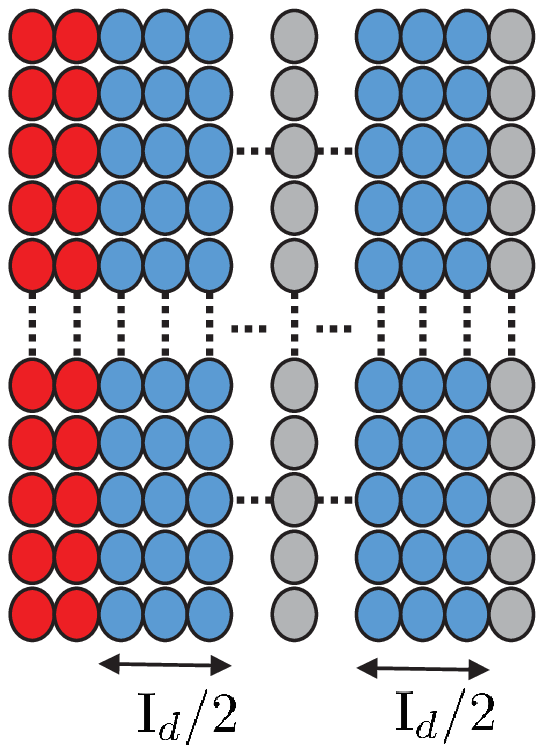}
    \caption{High mobility.}
    \label{fig:Full-MM}
  \end{subfigure}\hfill
  \begin{subfigure}[t]{0.25\textwidth} \centering
    \includegraphics[height=0.8\linewidth]{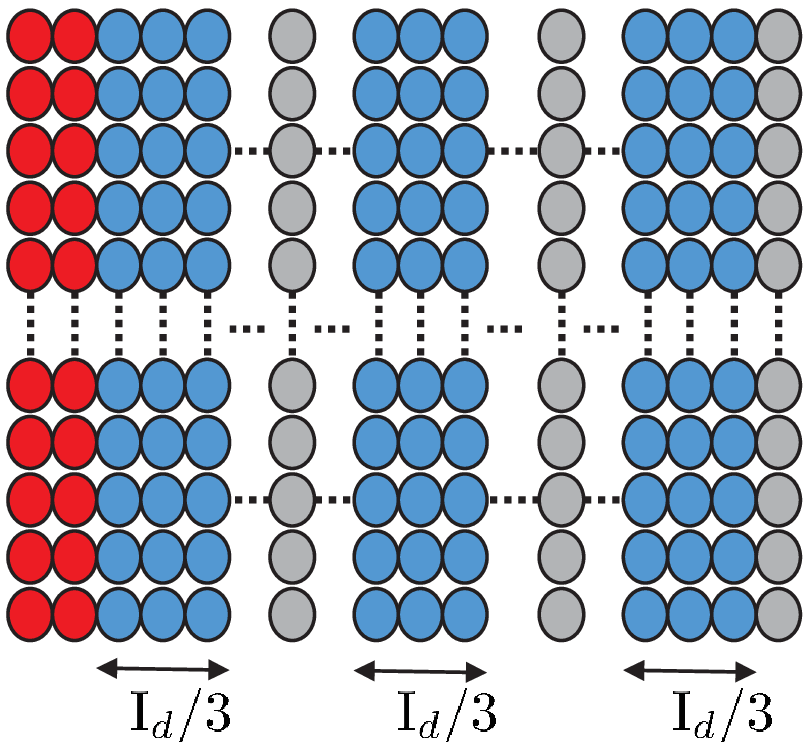}
    \caption{Very high mobility.}
    \label{fig:Full-HM}
  \end{subfigure}\hfill
  \begin{subfigure}[t]{0.25\textwidth} \centering
    \includegraphics[height=0.8\linewidth]{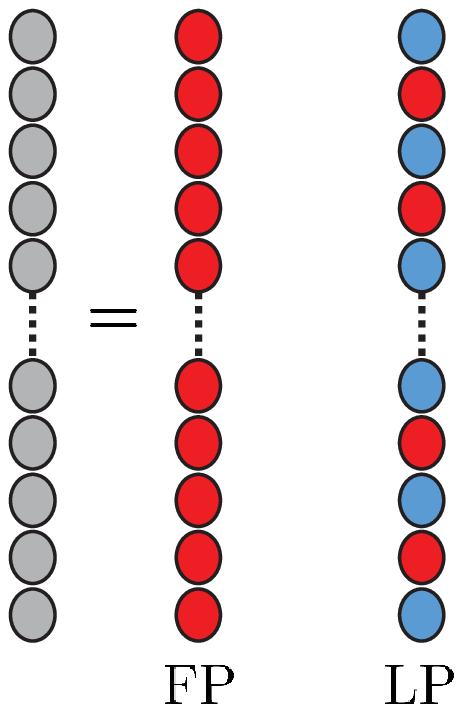}
    \caption{Pilot allocation schemes.}
    \label{fig:P_Conf}
  \end{subfigure}
  \caption{Proposed IEEE 802.11p frame structure employing different pilot allocation schemes.} 
  \label{fig:Full-P}
\end{figure*}

It is worth mentioning that the {\ac{DD}} estimation suffers from a considerable accumulated demapping error that is enlarged within the frame, especially at low \ac{SNR} region. Therefore, to reduce its impact, this error  $\hat{\tilde{\ma{h}}}_{\text{DD}_{i}}[k]$ is converted to $\hat{{\ma{h}}}_{\text{DD}_{i}}$ in the time domain using inverse fast Fourier transformation (IFFT), such that
\begin{equation}
\hat{{\ma{h}}}_{\text{DD}_{i}} = {\ma{F}}_K^{\text{H}} \hat{\tilde{\ma{h}}}_{\text{DD}_{i}},
\label{eq:IFFT}
\end{equation}
where ${\ma{F}}_K \in \compl^{K \times K}$ denotes the $K$-DFT matrix. Then, by truncating $\hat{{\ma{h}}}_{\text{DD}_{i}}$ to the significant $L$ channel taps, i.e.  
\begin{equation}
	\hat{{\ma{h}}}_{\text{DD}_{i,L}} =  \hat{{\ma{h}}}_{\text{DD}_{i}} ( 1:L).
	\label{eq:IFFT_TT}
\end{equation}
After that, $\hat{{\ma{h}}}_{\text{DD}_{i,L}}$ is converted back to the frequency domain such that
\begin{equation}
\hat{\tilde{\ma{h}}}_{\text{TT}_{i}} = {\ma{F}}_{\text{on}} \hat{{\ma{h}}}_{\text{DD}_{i,L}},
\label{eq:TT}
\end{equation}
where $ {\ma{F}}_{\text{on}} \in \compl^{K_{\text{on}} \times L}$ represents the scaled DFT matrix that is used to convert $\hat{{\ma{h}}}_{\text{DD}_{i,L}}$ back to frequency domain, which is obtained by selecting $\Kon$ rows, and $L$ columns from the $K$-DFT matrix. Applying the average time truncation operation to $\hat{\tilde{\ma{h}}}_{\text{DD}_{i}}[k]$ alleviates the effect of noise and enlarged demapping error. Moreover, 
$\hat{\tilde{\ma{h}}}_{\text{TT}_{i}}[k]$ estimated channel is further improved by applying frequency and time domain averaging consecutively as follows
\begin{equation}
\hat{\tilde{\ma{h}}}_{\text{FTT}_{i}}[k] = \sum_{\lambda = -\beta}^{\lambda = \beta} \omega_{\lambda} \hat{\tilde{\ma{h}}}_{\text{TT}_{i}}[k + \lambda], ~ \omega_{\lambda} = \frac{1}{2\beta+1}.
\label{eq: STA_4}
\end{equation}
The final {\ac{ADD-TT}} channel estimates is updated using time averaging between the previously {\ac{ADD-TT}} estimated channel and the frequency averaged channel in~{\eqref{eq: STA_4}}, such that
\begin{equation}
\hat{\tilde{\ma{h}}}_{\text{ADD-TT}_{i}}[k] = (1 - {\alpha})  \hat{\tilde{\ma{h}}}_{\text{ADD-TT}_{i-1}}[k] + {\alpha}\hat{\tilde{\ma{h}}}_{\text{FTT}_{i}}[k].
\label{eq: STA_5}
\end{equation}
Motivated by the fact that the vehicular time-variant channel can be modeled as a time-series forecasting problem, where historical data can be used to predict future observations~\cite{ref_timeseriers}.
The authors in~\cite{ref_TS_ChannelNet} apply \ac{SR-ConvLSTM} network on top of the {\ac{ADD-TT}} interpolation, where convolutional layers are added to the LSTM network in order to capture more vehicular channel features, hence improving the estimation performance. Therefore, the {\ac{ADD-TT}} estimated channel for the whole received frame is modeled as a low resolution image, and then \ac{SR-ConvLSTM} network is used after the {\ac{ADD-TT}} interpolation. Unlike {\ac{ChannelNet}} estimator where two {\acp{CNN}} are employed, {\ac{TS-ChannelNet}} estimator uses only {\ac{SR-ConvLSTM}} network, thus the overall computational complexity is relatively decreased. However, {\ac{TS-ChannelNet}} still suffers from high computational complexity due to integrating both LSTM and {\ac{CNN}} in one network.

%% file: Proposed_2DWI.tex
\section{Proposed Weighted Interpolation Estimators} \label{WI}

\label{DFT_Estimators}
The proposed {\ac{WI}} estimators employ different pilot allocation schemes to the IEEE 802.11p frame structure as shown in Fig.~\ref{fig:Full-P}. These frame structures are proposed motivated by the fact the vehicle velocity is a known parameter that can be exchanged   between all vehicular network nodes. For example, in urban environments (inside cities) the car velocity must not exceed 40 Kmphr, thus vehicles and road side units within this environment will use the low mobility frame structure, and similarly for highways environment. Therefore, the frame structure selection is performed in an adaptive manner according to the vehicle velocity. The proposed frame structures preserve the first two {\ac{LTS}} preamble symbols similar to IEEE 802.11p standard, so that the synchronization and frame detection processes will not be affected by the proposed modifications. Moreover, only $P$ pilot {\ac{OFDM}} symbols are required in the transmitted frame, such that $\tilde{\ma{Y}}_{P} = [\tilde{\ma{y}}^{(p)}_{1}, \dots,  \tilde{\ma{y}}^{(p)}_{q}, \dots, \tilde{\ma{y}}^{(p)}_{P}] \in \compl ^{K_{\text{on}}\times P}$. $q$ denotes the inserted pilot symbol index, where $1 \leq q \leq P$. The other $I_{d} = I - P$ {\ac{OFDM}} data symbols are preserved for actual data transmission.

The proposed estimators proceed according to two criteria: (\textit{i}) Vehicular mobility condition, where three frame structures are defined as shown in Fig.~\ref{fig:Full-P}. (\textit{ii}) The pilot allocation schemes, where there exist two schemes as shown in~\ref{fig:P_Conf}, such that
\begin{itemize}
    \item \textbf{Full pilot allocation ($\text{FP}$)}: where $K_\text{on}$ pilots are inserted within each pilot symbol and {\ac{LS}} estimation is applied to estimate the channel for each inserted pilot symbol. 
          The IEEE 802.11p basic {\ac{LS}} denoted as \emph{{\ac{SLS}}} estimation is applied using the two \ac{LTS} received preambles $\tilde{\ma{y}}_{{\text{LTS}}_{1}}[k]$, and $\tilde{\ma{y}}_{{\text{LTS}}_{2}}[k]$ as shown in~\eqref{eq: LS1}, and using each received pilot symbol such that
        \begin{equation}
        \hat{\tilde{\ma{h}}}_{{\text{SLS}}_{q}}[k] = \frac{\tilde{\ma{y}}^{(p)}_{q}[k]}{\tilde{\ma{p}}[k]} = \tilde{\ma{h}}_q[k] + \tilde{\ma{v}}_{{q}}[k], ~ k \in \Kon.
        \label{eq: SLSP}
        \end{equation}
        $\tilde{\ma{v}}_{q}[k]$ represents the noise at the $q$-th received pilot symbol. 
        On the other hand, the  \emph{{\ac{ALS}}} relies on the fact that $\tilde{\ma{h}}_{{q}} =\ma{F}_{\text{on}} {\ma{h}}_{q,L}$, where ${\ma{h}}_{q,L} \in \compl ^{L\times 1}$ denotes the channel impulse response at the $q$-th received pilot symbol that can be estimated as follows
        \begin{equation}
        \hat{\ma{h}}_{q,L} = \ma{F}_{\text{on}}^{\dagger} \hat{\tilde{\ma{h}}}_{{\text{LS}}_{q}} = {\ma{h}}_{q,L} +\ma{F}_{\text{on}}^{\dagger} \tilde{\ma{v}}_{q},~  k \in \Kon,
        \label{eq:ALS1}
        \end{equation}
        where $\ma{F}_{\text{on}}^{\dagger} = [(\ma{F}_{\text{on}}^{\text{H}} \ma{F}_{\text{on}})^{-1} \ma{F}_{\text{on}}^{\text{H}}]$ is the pseudo inverse matrix of $\ma{F}_{\text{on}}$. After that, the accurate {\ac{LS}} estimation can be obtained by applying the {\ac{DFT}} interpolation of $\hat{\ma{h}}_{q,L}$ as follows
        \begin{equation}
        \hat{\tilde{\ma{h}}}_{{\text{ALS}}_{q}} = \ma{F}_{\text{on}} \hat{\ma{h}}_{q,L}
        = \ma{F}_{\text{on}} {\ma{h}}_{q,L}  + \ma{W}_{\text{ALS}} \tilde{\ma{v}}_{{q}},  ~ k \in \Kon.
        \label{eq:ALSP}
        \end{equation}
        $\ma{W}_{\text{ALS}} = \ma{F}_\text{on} \ma{F}_\text{\text{on}}^{\dagger}$ denotes the accurate {\ac{LS}} {\ac{DFT}} interpolation matrix.
  
    \item \textbf{$L$ pilot allocation ($\text{LP}$)}: this allocation is motivated by the fact that vehicular channel models mainly consists of $12$ taps, then calculating  ${\ma{h}}_{q,L}$ requires only $L$ pilots in each pilot symbol, and can be computed as follows
    \begin{equation}
    \hat{\ma{h}}_{q,L} = \ma{F}_{p}^{\dagger} \hat{\tilde{\ma{h}}}_{{\text{LS}}_{q}} = {\ma{h}}_{q,L} +\ma{F}_{p}^{\dagger} \tilde{\ma{v}}_{q},~  k \in \Kp.
    \label{eq:dft4}
    \end{equation}
    $\hat{\tilde{\ma{h}}}_{\text{LS}_{q}}$ here represents the {\ac{LS}} estimated channel for the $q$-th pilot symbol at the $L$ equally spaced inserted pilot subcarriers, where $K_{p} = L$ and $\ma{F}_{p}^{\dagger} = [(\ma{F}_{p}^{\text{H}} \ma{F}_{p})^{-1} \ma{F}_{p}^{\text{H}}]$ is the pseudo inverse matrix of $\ma{F}_{p} \in \compl^{K_{\text{p}} \times L}$ which refers to the truncated DFT matrices obtained by selecting $\Kp$ rows, and $L$ columns from the $K$-DFT matrix. After that, {\ac{DFT}} interpolation based estimation can be used employing $\hat{\ma{h}}_{q,L}$ to estimate the channel for the $q$-th pilot symbol as follows
    \begin{eqnarray}
    \hat{\tilde{\ma{h}}}_{\text{DFT}_{q}} = \ma{F}_{\text{on}} \hat{\ma{h}}_{q,L} \nonumber 
    = \ma{F}_{\text{on}} {\ma{h}}_{q,L}  + \ma{W}_{\text{DFT}} \tilde{\ma{v}}_{{q}},  ~ k \in \Kon.
    \label{eq:dft5}
    \end{eqnarray}
Here $\ma{W}_{\text{DFT}} = \ma{F}_\text{on} \ma{F}_\text{p}^{\dagger}$ is the DFT interpolation matrix.
\end{itemize}

\begin{figure*}[!b]
\begin{equation}
\small
\begin{split}
 \ma{C}_{{f}} &= \Ex{\tilde{\ma{H}}_{{f}} \hat{\tilde{\ma{H}}}^{H}_{{f}}} \left[\Ex{\hat{\tilde{\ma{H}}}_{{f}} \hat{\tilde{\ma{H}}}^{H}_{{f}}} \right]^{-1} = \begin{bmatrix} \Ex{ \tilde{\ma{H}}_{{f}} \hat{\tilde{\ma{h}}}^{H}_{{q}}} & \Ex{ \tilde{\ma{H}}_{{i}} \hat{\tilde{\ma{h}}}^{H}_{{q+1}}} \end{bmatrix} \begin{bmatrix}
 \Ex{\norm{{\tilde{\ma{h}}}_{{q}}}^2} + E_{{{q}}} & \Ex{{\tilde{\ma{h}}}_{{q}} {\tilde{\ma{h}}}^{H}_{{q+1}}}\\
 \Ex{{\tilde{\ma{h}}}_{{q+1}} {\tilde{\ma{h}}}^{H}_{{q}}} & \Ex{\norm{{\tilde{\ma{h}}}_{{q+1}}}^2} + E_{{{q+1}}}
 \end{bmatrix}^{-1} \\
 &=    \begin{bmatrix}
 J_{0} (2 \pi f_{\text{d}} (f-1) T_{\text{s}}) &
 J_{0} (2 \pi f_{\text{d}} (I_{f} + 1 - f) T_{\text{s}})
 \end{bmatrix}  \begin{bmatrix}
 1 + E_{{{\Phi}_{q}}} & J_{0} (2 \pi f_{\text{d}} I_{f} T_{\text{s}})\\
 J_{0} (2 \pi f_{\text{d}} I_{f} T_{\text{s}}) & 1 + E_{{{q+1}}}
 \end{bmatrix}^{-1}.
\end{split}
\label{eq:ci}
\end{equation}
\end{figure*}

After estimating the channel at the pilot symbols according to the selected configuration, the proposed {\ac{WI}} estimators apply the following two steps
\begin{itemize}
      \item \textbf{Grouping}: The estimated channels of the $P$ pilot symbols are grouped into $P$ matrices, such that
      
      \begin{equation}
          \hat{\tilde{\ma{H}}}_{q} = [\hat{\tilde{\ma{h}}}_{q-1}, \hat{\tilde{\ma{h}}}_{q}],~ q = 1, \cdots P.
      \end{equation}
      
      We note that $\hat{\tilde{\ma{h}}}_{0} = \hat{\tilde{\ma{h}}}_{\text{LTS}}$ and $\hat{\tilde{\ma{h}}}_{q}$ refers to the implemented {\ac{LS}} estimation according to the utilized pilot allocation scheme.
      
      \item \textbf{Weighted Interpolation estimation}:  According to the grouped $\hat{\tilde{\ma{H}}}$ matrices, the received frame can be divided into $P$ sub-frames, where $f$ denotes the sub-frame index, such that $1 \leq f \leq P$. Therefore, the estimated channel for the $i$-th received {\ac{OFDM}} symbol within each $f$-th sub-frame can be expressed as follows
      \begin{equation}
          \hat{\tilde{\ma{H}}}_{{\text{WI}}_{f}} =  \hat{\tilde{\ma{H}}}_{{f}} \ma{C}_{f}
      \label{eq:WI_LS}
      \end{equation}
      where $\hat{\tilde{\ma{H}}}_{f} \in \compl^{ K_{\text{on}} \times 2} $ denotes {\ac{LS}} estimated channels at the pilot symbols within the $f$-th sub-frame. $\ma{C}_{f} \in \mathbb{R}^{2 \times I_{f}}$ denotes the interpolation weights of the $I_{f}$ {\ac{OFDM}} data symbols within the $f$-th sub-frame. We note that $I_{f}$ is calculated according to the selected frame structure where it is equal to $I_{d}$, $\frac{I_{d}}{2}$, and $\frac{I_{d}}{3}$ in low, high, and very high mobility scenarios respectively. 
      
      Therefore, the estimated channel for the received {\ac{OFDM}} data symbols can be seen as a weighted summation of the $\hat{\tilde{\ma{H}}}_{f}$ matrix. $\ma{C}_{f}$ interpolation weights are calculated by minimizing the {\ac{MSE}} between the ideal channel $\tilde{\ma{H}}_{{f}}$, and the {\ac{LS}} estimated channel at the {\ac{OFDM}} pilot symbols $\hat{\tilde{\ma{H}}}_{{f}}$ as derived in~\cite{ref_interpolation_matrix}. This minimization results in $\ma{C}_{{f}}$ expressed in~\eqref{eq:ci},  where $J_{0}(.)$ is the zeroth order Bessel function of the first kind, $T_{\text{s}}$ is the received {\ac{OFDM}} data symbol duration, and $E_{{{q}}}$ denotes the overall noise of the estimated channel at the $q$-th pilot symbol. $E_{{{q}}}$ is calculated according to the chosen pilot allocation scheme, and employed {\ac{LS}} estimation, where it equals to $\sigma^2$, $\sigma^2  \trace{\ma{W}_{\text{ALS}} \ma{W}_{\text{ALS}}^{\text{H}}}$, and $\sigma^2  \trace{\ma{W}_{\text{DFT}} \ma{W}_{\text{DFT}}^{\text{H}}}$ for {\ac{SLS}}, {\ac{ALS}} and LP {\ac{LS}} estimation respectively.
\end{itemize}

In fact, the weight elements of $\ma{C}_{{f}}$ for all the sub-frames can be calculated offline for several vehicular scenarios by employing different $f_{\text{d}}$ and $I_{d}$ values, Therefore, decreasing the online complexity and making the proposed {\ac{WI}} estimators more efficient in real cases scenarios. Moreover, a trade-off between the mobility condition controlled by $f_{\text{d}}$, the inserted pilot symbols $P$, and the used frame length $I_{d}$ should be considered. As the mobility increases, more pilot symbols should be inserted within the transmitted frame. Therefore, this trade-off is managed by the vehicular application requirements, so that the transmitter will adapt the transmission parameters according to these requirements. Fig.~\ref{fig:soa_Figure} shows the proposed {\ac{WI}} estimators block diagram.

%% file: CNN.tex
\section{CNN-based channel estimation}\label{CNN}
\begin{figure*}[t]
	\centering
	\includegraphics[width=2\columnwidth]{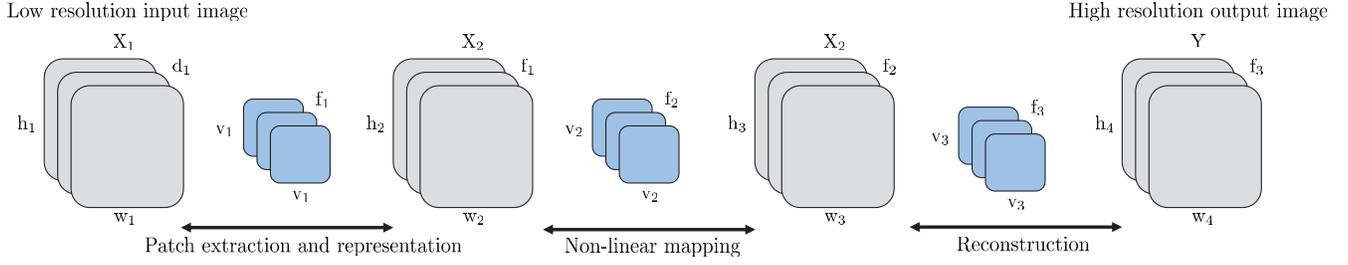}
	\caption{SR-CNN architecture.}
	\label{fig:sr_cnn}
\end{figure*}

\begin{figure*}[t]
	\centering
	\includegraphics[width=2\columnwidth]{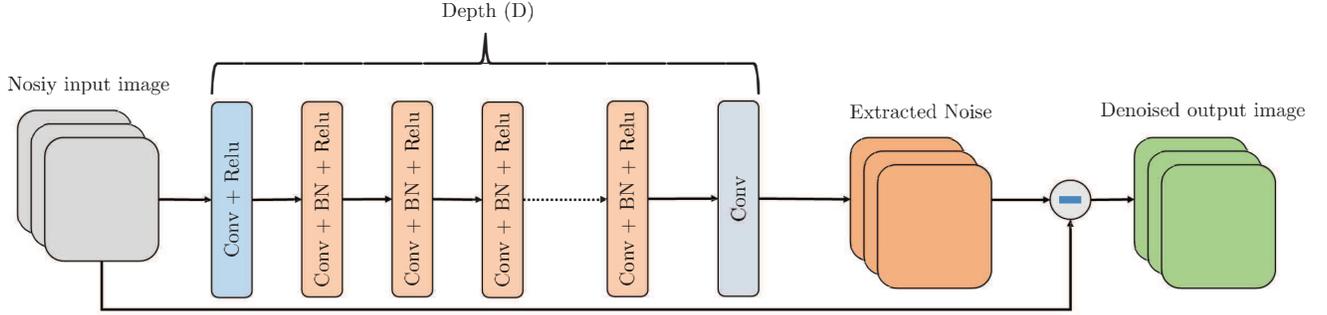}
	\caption{DN-CNN architecture.}
	\label{fig:dn_cnn}
\end{figure*}
\subsection{{\ac{CNN}} Overview}
A {\ac{CNN}} is a neural network that consists of one or more layers and is used mainly for processing structured arrays of data such as images~\cite{ref_CNN}. {\ac{CNN}} network has generally become the state of the art for many visual applications such as image classification, due to its great ability in extracting patterns from the input image. {\ac{CNN}} is a set of several layers stacked together in order to accomplish the required task. These layers include
\begin{itemize}
    \item Input layer: It represents the 2D or 3D input image.
    \item Convolutional layer: It is used for feature extraction that is implemented using predefined filters called kernels. The kernel scans the input matrix to fill the output matrix denoted as feature map. 
    \item Pooling layer: This layer is applied to reduce the number of parameters when the images are too large. Pooling operation is also called sub-sampling or down-sampling which reduces the dimensionality of each feature map but retains important information. 
    \item Activation layer: An activation function is applied to the {\ac{CNN}} layer's output. The most used common activation function is the ReLU function. It is used to introduce non-linear processing to the {\ac{CNN}} architecture since the input-output {\ac{CNN}} pairs relation might be non-linear. 
    \item Fully connected layer: This layer forms the last block of the {\ac{CNN}} architecture, and it is utilized mainly in classification problems. It is a simple feed forward neural network layer, consisting of one or more hidden layers.
    \item Batch normalization: It is used to make the {\ac{CNN}} output more stable by normalizing the output of each layer. Moreover, batch normalisation layer reduces overfitting and speeds up the {\ac{CNN}} training.
    \item Output layer: This layer is configured according to the studied problem, for example, in classification problems the {\ac{CNN}} output layer is a fully connected layer with softmax activation function, while in regression problems, the {\ac{CNN}} output does not use any activation function.
\end{itemize}

Automatically detecting meaningful features given only an image and its label is not a trivial task. Therefore, the convolutional layers are the most essential elements within any {\ac{CNN}} since the convolutional layers learn such complex features by updating their kernels values during training phase to obtain the best input-output mapping. This update is performed by minimizing the {\ac{CNN}} loss function that measures how far are the inputs from the outputs. After that, the performance of the trained {\ac{CNN}} model is evaluated in the testing phase where new unobserved inputs are fed to the trained {\ac{CNN}} model.

\subsection{{\ac{SR-CNN}} and {\ac{DN-CNN}} Networks}

Besides classifications problems, there are  special {\ac{CNN}} architectures denoted as {\ac{SR-CNN}} and {\ac{DN-CNN}} that are mainly used for regression problems. {\ac{SR-CNN}}~\cite{ref_SRCNN}  is used to improve the quality of the input image, where the input-output mapping is represented as a deep {\ac{CNN}} that takes the low-resolution image as the input and outputs the high-resolution one. {\ac{SR-CNN}} consists of three convolutional layers as shown in \figref{fig:sr_cnn}: ($\textit{i}$) Patch extraction and representation, ($\textit{ii}$) Non-linear mapping, and ($\textit{iii}$) Reconstruction. Let {$\ma{X}_{m} \in \mathbb{R}^{h_{m} \times w_{m} \times d_{m}}$}  be the input image to the $m$-th {\ac{SR-CNN}} convolutional layer, where $h_{m}$, $w_{m}$, and $d_{m}$ denote the image height, width, and depth respectively, and {$\ma{X}_{1} \in \mathbb{R}^{h_{1} \times w_{1} \times d_{1}}$} represents the input image. The convolutional filters are denoted by the weights and biases matrices $\ma{W}_{m} \in \mathbb{R}^{v_{m} \times v_{m} \times d_{m} \times f_{m}}$ and $\ma{B}_{m} \in \mathbb{R}^{f_{m} \times 1}$, where $v_{m}$ refers to the convolutional filter size, and $f_{m}$ denotes the total number of filters used in the {\ac{SR-CNN}} $m$-th convolutional layer. The $m$-th convolution layer applies $f_{m}$ convolution operation to its $\ma{X}_{m}$ input image as follows
\begin{equation}
\ma{Y}_{(m)} = \ma{F}_{(m)} \Big( {\ma{W}_{(m)}} * \ma{X}_{(m)} \Big),~\ma{X}_{(m+1)} = \ma{Y}_{(m)},
\label{eq:cnn_layer_output}
\end{equation}
where $\ma{F}_{(m)}$ represents the activation function applied to the $m$-th convolutional layer output. We note that, it is possible to add more convolutional layers to increase the non-linearity mapping between the {\ac{SR-CNN}} convolutional layers, but this will increase the complexity of the {\ac{SR-CNN}} model and thus demands more training time. The integrated {\ac{SR-CNN}} architecture related mainly to the investigated problem and the dataset structure, whereas the best {\ac{SR-CNN}} architecture and parameters can be fine tuned by  intensive experiments or by using  the grid search
algorithm~\cite{ref_grid_search} that selects the best suitable {\ac{SR-CNN}} hyper parameters in terms of both performance and complexity. The loss function of the {\ac{SR-CNN}} is represented by the \ac{MSE} between the  reconstructed high resolution images $\ma{Y}$, and the true images ${{\ma{Y}}}^{(T)}$, such that

\begin{equation}
\text{MSE}_{\text{SR-CNN}} = \frac{1}{N_{\text{Train}}} \sum_{i = 1}^{N_{\text{Train}}} \left\| {\ma{Y}}_i - {{\ma{Y}_{i}}^{(T)}} \right\|^2,
\label{eq:mse_srcnn}
\end{equation}

\noindent where $N_{\text{Train}}$ denotes the number of training samples. The minimization of $\text{MSE}_{\text{SR-CNN}}$ is achieved by updating $\ma{W}$ and $\ma{B}$ matrices such that the best input-output mapping is obtained. To do so, various optimizers can be exploited such that the stochastic gradient descent~\cite{ref_SGD}, root mean square prop~\cite{ref_RMSP}, and adaptive moment estimation (ADAM)~\cite{ref_ADAM}.

On the other hand, {\ac{DN-CNN}~\cite{ref_DNCNN} improves the image quality by separating the noise from the input noisy image using a special {\ac{CNN}} architecture. After that, the input noisy image is subtracted from the extracted noise resulting in the denoised image. In order to extract the noise from a noisy image, residual learning~\cite{ref_residual_learning} is applied so that the noise included in the input image is learnt in the {\ac{DN-CNN}} training phase by minimizing the following {\ac{DN-CNN}} loss function

\begin{table}[t]
	\renewcommand{\arraystretch}{1.3}
	\centering
	\small
	\caption{Optimized {\ac{SR-CNN}} and {\ac{DN-CNN}} parameters.}
	\label{tb:CNN_params}
	\begin{tabular}{|c|c|}
		\hline
		\textbf{Parameter}      & \textbf{Values}                    \\ \hline
		Input/Output dimensions & $ 2K_{\text{on}} \times I \times 1$   \\ \hline
		SRCNN ($v_{1},~f_{1};~v_{2}, f_{2};~v_{3},~f_{3}$) & (9,32;~1,16;~ 5,1)   \\ \hline
		DNCNN ($v,~f$) & (3, 16)   \\ \hline
		DNCNN D & 7  \\ \hline
		Activation function              & ReLU                      \\ \hline
		Number of epochs        & 250                                \\ \hline
		Training samples        & 8000                             \\ \hline
		Testing samples        & 2000                             \\ \hline
		Batch size          & 128                                    \\ \hline
		Optimizer       & ADAM                                       \\ \hline
		Loss function      & MSE                                     \\ \hline
		Learning rate        & 0.001                                 \\ \hline
		Training SNR        & 30 dB                                 \\ \hline
	\end{tabular}
\end{table}

\begin{equation}
\text{MSE}_{\text{DN-CNN}} = \frac{1}{N_{\text{Train}}} \sum_{i = 1}^{N_{\text{Train}}} \left\| {{\ma{V}}_i} - {{\ma{V}_{i}}^{(T)}} \right\|^2,
\label{eq:mse_dncnn}
\end{equation}
\noindent where ${{\ma{V}}_i}$ represents the extracted noise included in the noisy input image $\ma{X}_{i}$, and ${{\ma{V}_{i}}^{(T)}}$ denotes the exact noise, such that ${\ma{V}_{i}}^{(T)} = {{\ma{Y}}_i}^{(N)} - \ma{X}_{i}$. As shown in \figref{fig:dn_cnn}, the {\ac{DN-CNN}} employs in the first layer convolution and ReLU activation function to generate the initial feature maps from the input noisy image. After that, successive (Conv + BN + ReLU) layers are considered to extract the noise features, since the analysis performed in~\cite{ref_DNCNN} shows that integrating convolution with batch normalization followed by ReLU, can gradually separate the clean image structure from the noisy observation through the {\ac{DN-CNN}} hidden layers. Finally, a convolutional layer is used for output image reconstruction. In summary, the {\ac{DN-CNN}} architecture has two main tasks: the residual learning is used to learn the noise features, and the batch normalization which is incorporated to speed up training as well as to boost the denoising performance. The main complexity of the {\ac{DN-CNN}} architecture lies in its depth ($\text{D}$) that denotes the number of used (Conv+ BN + ReLU) layers. As $\text{D}$ increases the complexity of the {\ac{DN-CNN}} architecture increases. Similarly to {\ac{SR-CNN}}, all the {\ac{DN-CNN}} parameters can be tuned using grid search algorithm. 

Unlike ChannelNet estimator, where both {\ac{SR-CNN}} and {\ac{DN-CNN}} networks are used on top of the \ac{RBF} interpolation, extensive experiments are applied in this paper using the hyper parameters tuning grid search algorithm~\cite{ref_grid_search} in order to select the best {\ac{CNN}} network configuration corresponding to the mobility condition. Based on the selected CNN parameters, an optimized {\ac{SR-CNN}} is employed on top of the {\ac{WI}} estimators in a low mobility scenario, whereas an optimized {\ac{DN-CNN}} is considered in high mobility one. By doing so, better performance can be achieved with a significant decrease in the overall computational complexity as we discuss in the next sections. We note that employing, low-complexity {\ac{SR-CNN}} and {\ac{DN-CNN}} architectures is due to the good performance of the initial {\ac{WI}} estimation, since as the initial estimation accuracy increases, the CNN computational complexity decreases. Moreover, our investigations show that by employing the proposed {\ac{WI}} estimators, there is no need to use both {\ac{SR-CNN}} and {\ac{DN-CNN}}, since the computational complexity  increases without any significant performance gain.

We note that {\acp{CNN}} networks work with real valued data, therefore after applying the {\ac{WI}} interpolation estimator, $\hat{\tilde{\ma{H}}}_{{\text{WI}}} \in \compl^{K_{\text{on}} \times I_{d}}$ is converted from complex to real valued domain by stacking the real and imaginary values vertically in one matrix, such that $\hat{\tilde{\ma{H}}}^{(R)}_{{\text{WI}}} \in \mathbb{R}^{2K_{\text{on}} \times I_{d}}$.
Then $\hat{\tilde{\ma{H}}}^{(R)}_{{\text{WI}}}$ is fed as an input to the optimized {\ac{SR-CNN}} or {\ac{DN-CNN}} according to the mobility scenario. Finally, the output of the employed network is converted back to the complex domain. The optimized {\ac{SR-CNN}} and {\ac{DN-CNN}} networks are trained on {\ac{SNR}}$ = 30~\text{dB}$ for each mobility scenario, since in high {\ac{SNR}} region, the {\ac{CNN}} network is able to learn better the channel due to the low noise impact in high {\ac{SNR}} region~\cite{ref_LS_DNN}. Moreover, ADAM optimizer is used with {\ac{MSE}} loss function. \tabref{tb:CNN_params} shows the proposed optimized {\ac{SR-CNN}} and {\ac{DN-CNN}} parameters.

%% file: Simulation_Results_Updated.tex
\begin{table*}[t!]
\renewcommand{\arraystretch}{1.3}
\small
\caption{Vehicular channel models characteristics following Jake's Doppler spectrum.}
\label{tb:VCMC}
\begin{tabular}{|c|c|c|c|c|c|}
\hline
\begin{tabular}{@{}c@{}}\textbf{Channel} \\\textbf{model} \end{tabular}  & \begin{tabular}{@{}c@{}}\textbf{Channel} \\\textbf{taps} \end{tabular} & \begin{tabular}{@{}c@{}}\textbf{Vehicle velocity} \\\textbf{[kmph]} \end{tabular} & \begin{tabular}{@{}c@{}}\textbf{Doppler} \\\textbf{shift [Hz]} \end{tabular}  & \textbf{Average path gains {[}dB{]}}                                                                                       & \textbf{Path delays {[}ns{]}}                                                                            \\ \hline
VTV-UC                 & 12                             & 45                                & 250                                 & \begin{tabular}[c]{@{}c@{}}{[}0, 0, -10, -10, -10, -17.8, -17.8,\\ -17.8, -21.1, -21.1, -26.3, -26.3{]}\end{tabular}     & \begin{tabular}[c]{@{}c@{}}{[}0, 1, 100, 101, 102, 200, 201,\\202, 300, 301, 400, 401{]}\end{tabular} \\ \hline
VTV-SDWW               & 12                             & 100-200                                  & 500-1000                                & \begin{tabular}[c]{@{}c@{}}{[}0, 0, -11.2, -11.2, -19, -21.9, -25.3,\\ -25.3, -24.4, -28, -26.1, -26.1{]}\end{tabular} & \begin{tabular}[c]{@{}c@{}}{[}0, 1, 100, 101, 200, 300, 400,\\401, 500, 600, 700, 701{]}\end{tabular} \\ \hline
\end{tabular}
\end{table*}
\section{Simulation Results} \label{results}
\begin{figure*}[ht!]
	\setlength{\abovecaptionskip}{3pt plus 3pt minus 2pt}
	\centering
	\includegraphics[width=2\columnwidth]{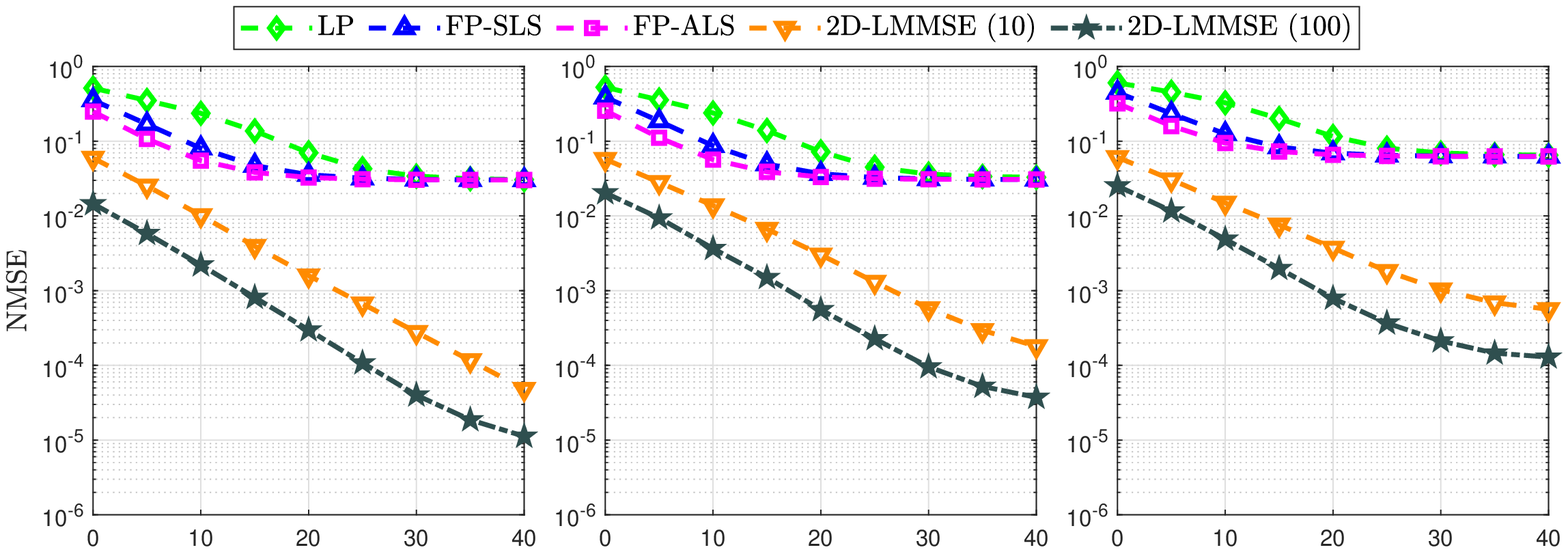}\\[-4ex]
	\subfloat[\label{fig:NMSE_WI} NMSE for linear methods.]{\hspace{.5\linewidth}} \\[-2ex]
	\vspace*{12pt}
	\includegraphics[width=2\columnwidth]{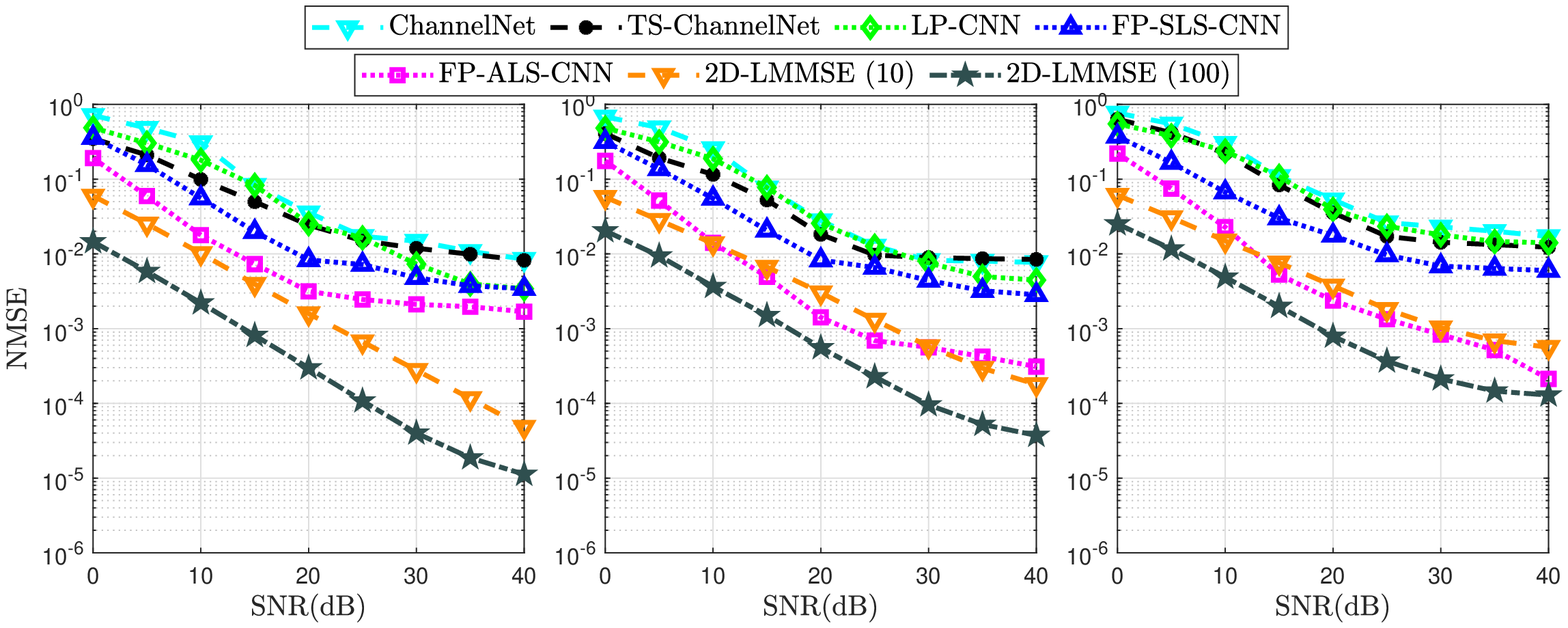}\\[-4ex]
	\subfloat[\label{NMSE_WI_CNN}  NMSE with DL post processing.]{\hspace{.5\linewidth}}
	\caption{NMSE employing $I = 100$, mobility from left to right: low  ($v = 45~\text{Kmph}, f_{d} = 250$ Hz), high  ($v = 100~\text{Kmph}, f_{d} = 500$ Hz), very high  ($v = 200~\text{Kmph}, f_{d} = 1000$ Hz). The CNN refers to SRCNN and DNCNN in low and high/very high) mobility scenarios, respectively.}
\end{figure*}
In this section, the performance of the proposed linear {\ac{WI}} estimators is evaluated compared to the conventional  2D {\ac{LMMSE}} that exploits all the pilots defined in the IEEE 802.11p standard, {\ac{ChannelNet}}, and {\ac{TS-ChannelNet}} estimators using {\ac{BER}} and {\ac{NMSE}}. The simulations are conducted employing three vehicular scenarios as shown in {\tabref{tb:VCMC}}. These scenarios are based on the {\ac{TDL}} vehicular channel models proposed in~\cite{ref_channel1}, which are obtained by a measurement campaign that was implemented in metropolitan Atlanta, and can be defined as follows

\begin{itemize}
    \item \textbf{Low mobility}: where VTV Urban Canyon (VTV-UC) vehicular channel model is considered. VTV-UC channel model has been measured between two vehicles moving in a dense urban traffic environment at ${V} = 45$ Kmph which is equivalent to  ${f}_{d} = 250$ Hz.
    
    \item \textbf{High mobility}: This scenario measures the communication channel between two vehicles moving on a highway having center wall between its lanes. Moreover, the vehicles are moving in the same direction at ${V} = 100$ Kmph which is equivalent to ${f}_{d} = 500$ Hz with a $300$–$400$ m separation distance between them. This vehicular channel model is denoted as VTV Expressway Same Direction with Wall (VTV-SDWW).     
    \item \textbf{Very high mobility}: In order to further evaluate the performance of the benchmarked channel estimators, VTV-SDWW vehicular channel model is used with ${V} = 200$ Kmph which implies ${f}_{d} = 1000$.
\end{itemize}
\begin{figure*}[t]
	\setlength{\abovecaptionskip}{3pt plus 3pt minus 2pt}
	\centering
	\includegraphics[width=2\columnwidth]{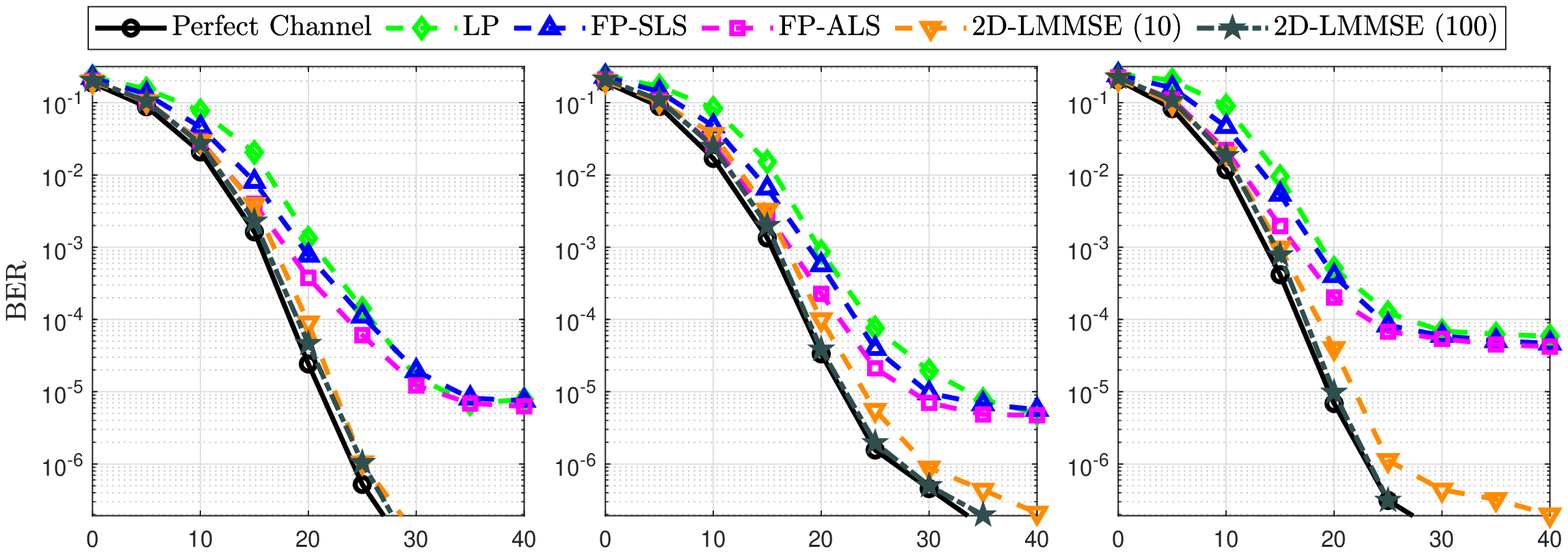}\\[-4ex]
	\subfloat[\label{BER_QPSK_WI} BER with linear estimation.]{\hspace{.5\linewidth}} \\[-2ex]
	\vspace*{12pt}
	\includegraphics[width=2\columnwidth]{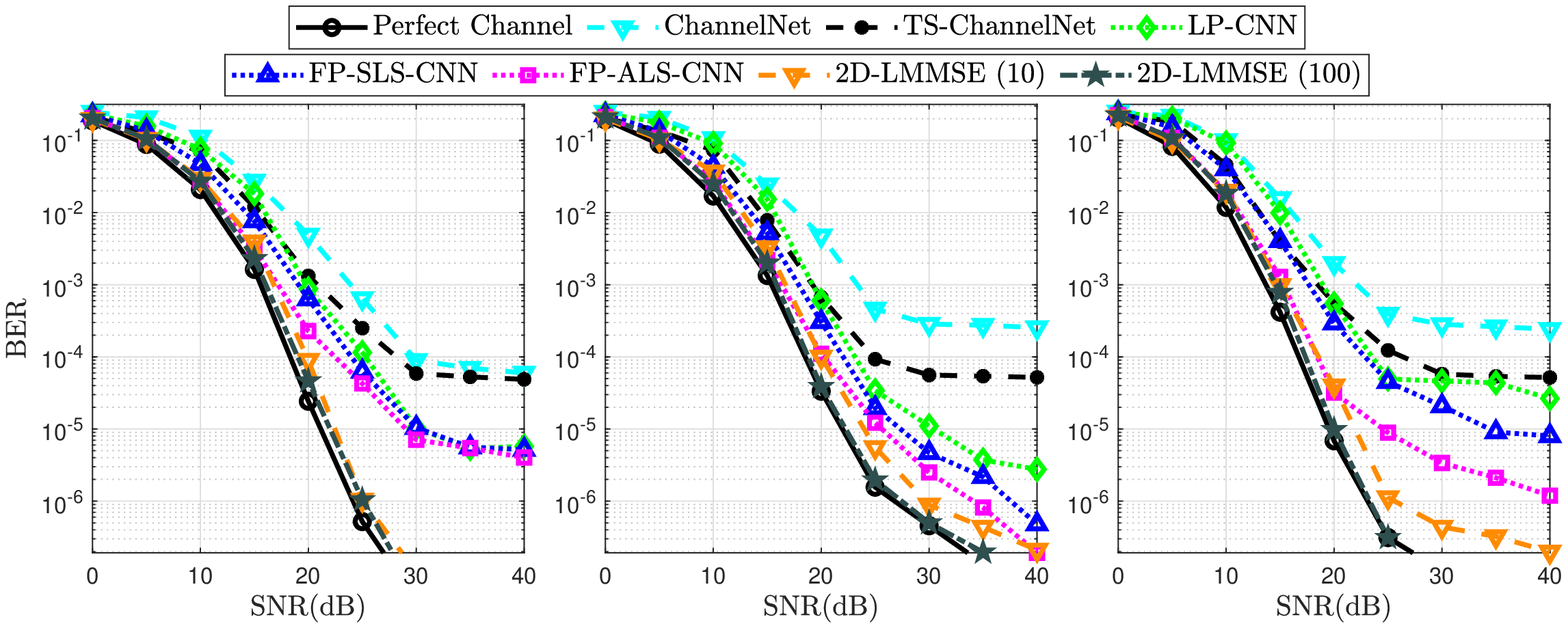} \\[-4ex]
	\subfloat[\label{BER_QPSK_WI_CNN} BER with DL post processing.]{\hspace{.5\linewidth}} \\
	\caption{BER for $I = 100$, QPSK, mobility from left to right: low  ($v = 45~\text{Kmph}, f_{d} = 250$ Hz), high  ($v = 100~\text{Kmph}, f_{d} = 500$ Hz), very high  ($v = 200~\text{Kmph}, f_{d} = 1000$ Hz). The CNN refers to SRCNN and DNCNN in low and high/very high) mobility scenarios, respectively.}
\end{figure*}
\subsection{NMSE Evaluation}
The {\ac{NMSE}} performance of the proposed estimators depends mainly on the employed {\ac{WI}} estimation, where using low number of pilots as the case in the proposed WI-LP estimator leads to considerable performance degradation as we can notice in \figref{fig:NMSE_WI}. Whereas, the proposed WI-FP-SLS and WI-FP-ALS achieve better performance than WI-LP due to the employment of full pilots symbols within the frame. The accuracy of  WI-FP-ALS is higher at low \acp{SNR} due to the exploitation of the frequency correlation via interpolation at the cost of increased complexity. By employing all the pilots in the frame ($4$ pilots per OFDM symbol), 2D-LMMSE (100) significantly outperforms the {\ac{WI}} estimators. In order to reduce the latency, we consider to apply the 2D-LMMSE on subframe basis with a subframe length of $10$ OFDM symbols denoted as 2D-LMMSE (10). The estimation accuracy in this case is reduced by $10$ dB with the decrease of the subframe length. Nevertheless, the accuracy of  2D-LMMSE (10) is  better than the WI estimators due to the high correlation of the pilots.
 
The performance of the {\ac{WI}} approaches is reduced by the increase of mobility. This mainly depends on the frame structure,  and the Doppler spread through the coherence interval between the pilots symbols, which can be defined as 
\begin{equation}
\Delta_C = \Delta_p I f_d T_{s},
\end{equation}
where $T_s$ is the \ac{OFDM} symbol duration, $\Delta_p I$ number of symbols between successive  pilot symbols. A smaller value indicates more correlation between the pilots that can be exploited in the time-domain interpolation.  
Based on the frame structure shown in   \figref{fig:Full-P}, at very high mobility ($f_d = 1000$ Hz), $\Delta_p = 33$ symbols, which corresponds to $\Delta_C = 33\times 10^3 T_s$, whereas in low and high mobility, $\Delta_C = 25 \times 10^3 T_s$. As a consequence, it is clear that the \ac{NMSE} at very high mobility increases compared to the other cases  because of the larger coherence interval between the pilots. The high mobility case is only influenced  by the Doppler interference, which can be observed from the error floor at high \acp{SNR}. This is also the situation in both 2D-LMMSE  estimators as the \ac{NMSE} increases with the increase of mobility at high \acp{SNR}.

\figref{NMSE_WI_CNN} {shows the NMSE performance of the CNN-based estimators. It can be noticed that the proposed estimators are able to outperform the ChannelNet and TS-ChannelNet estimators in different mobility scenarios. It is worth mentioning that using CNN post processing after the WI-based estimators reveals a considerable robustness against mobility. This is due to the ability of the optimized SR-CNN and DN-CNN in significantly alleviating the impact of Doppler interference. The DL-based post processing networks provide a performance trade-off between the linear WI and 2D-LMMSE using the full pilots in the frame.
\begin{figure*}[t]
	\setlength{\abovecaptionskip}{3pt plus 3pt minus 2pt}
	\centering
	\includegraphics[width=2\columnwidth]{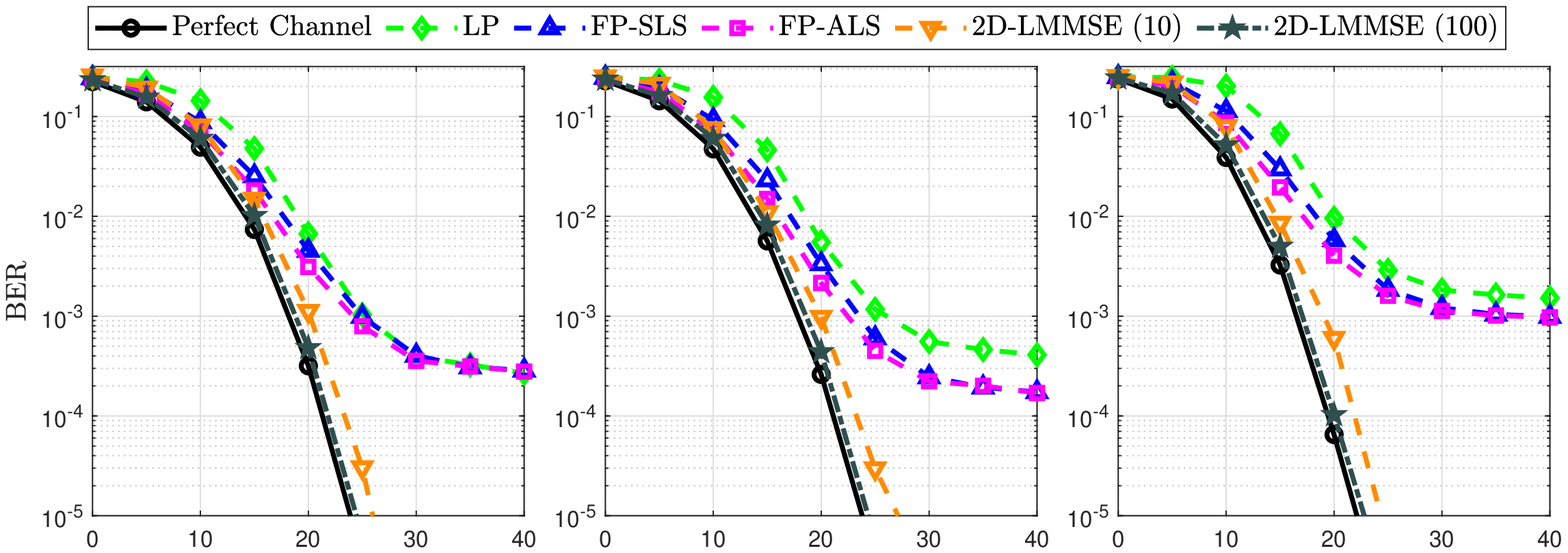}\\[-4ex]
	\subfloat[\label{BER_16QAM_WI} BER with linear estimation.]{\hspace{.5\linewidth}} \\[-2ex]
	\vspace*{12pt}
	\includegraphics[width=2\columnwidth]{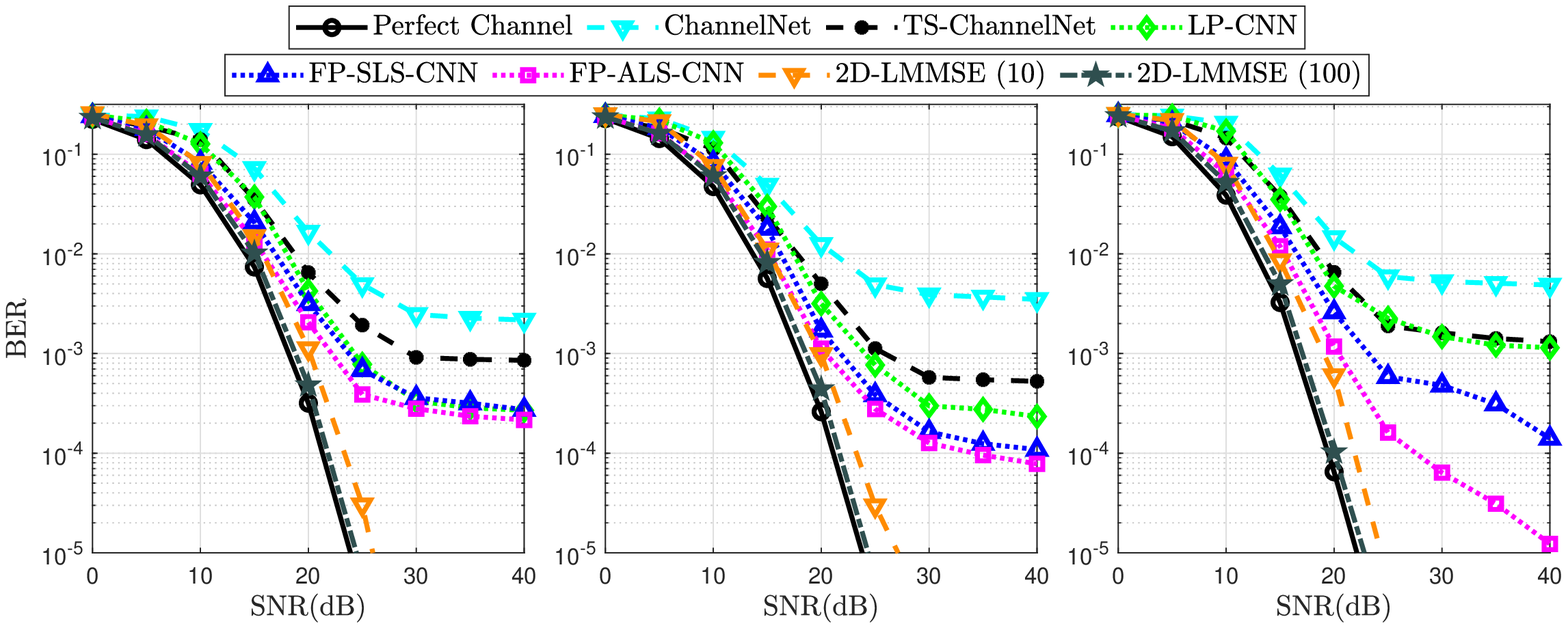} \\[-4ex]
	\subfloat[\label{BER_16QAM_WI_CNN} BER with DL post processing.]{\hspace{.5\linewidth}} \\
	\caption{BER for $I = 100$, 16QAM, mobility from left to right: low  ($v = 45~\text{Kmph}, f_{d} = 250$ Hz), high  ($v = 100~\text{Kmph}, f_{d} = 500$ Hz), very high  ($v = 200~\text{Kmph}, f_{d} = 1000$ Hz). The CNN refers to SRCNN and DNCNN in low and high/very high) mobility scenarios, respectively.}
\end{figure*}
\subsection{BER Evaluation}
\figref{BER_QPSK_WI} depicts the \ac{BER}  using the discussed linear estimation techniques employing QPSK. The relative  performance follows the same trend of the channel estimation performance shown in \figref{fig:NMSE_WI} in each mobility case. In general, the impact of the estimation error is lower in low \ac{SNR} region and this impact increases with the increase of the SNR. Although the \ac{NMSE} gap between WI-FP-ALS,  WI-FP-SLS, WI-LP  decreases with the increase of the \ac{SNR}, the \ac{BER} gain achieved by WI-FP-ALS is maintained until reaching an error floor. 
In different mobility scenarios, the trade-off between estimation error and time diversity gain can be observed where the performance is attributed to the main following factors; i) channel estimation error, and ii) time diversity due to increased Doppler spread. The estimation error degrades the \ac{BER} performance, whereas the time diversity improves it.
In total, the estimation error dominates over the  diversity gain. Nevertheless, the case of 2D-LMMSE experiences improvement due to better channel estimation, although the codeword is smaller than that used in WI-FP-ALS.

The impact of the proposed DL-based post processing is shown in \figref{BER_QPSK_WI_CNN}. First, it can be clearly seen that the post processing enhances the \ac{BER} as a result of enhancing the channel estimation, \figref{NMSE_WI_CNN}.  
Next, we compare our proposed linear,  and DL-enhanced estimation with the SoA DL-based  {\ac{ChannelNet}} and {\ac{TS-ChannelNet}}.
We can observe the significant {\ac{BER}} performance superiority of the proposed estimators, where {\ac{WI}}-LP records similar performance as {\ac{TS-ChannelNet}}, while {\ac{WI}}-FP-{\ac{SLS}} estimator slightly outperforms {\ac{WI}}-LP by around $1$ dB gain in terms of {\ac{SNR}} for a BER = $10^{-3}$. On the other hand, the proposed {\ac{WI}}-FP-{\ac{ALS}} estimator outperforms both {\ac{ChannelNet}} and {\ac{TS-ChannelNet}} by around $6$ dB and $3$ dB gain in terms of {\ac{SNR}} for a BER = $10^{-3}$, respectively. 

The performance of {\ac{ChannelNet}} and {\ac{TS-ChannelNet}} accounts of the predefined fixed parameters in the applied interpolation scheme, where the RBF interpolation function and the ADD-TT frequency and time averaging parameters need to be updated in a real-time manner. Moreover, the ADD-TT interpolation employs only the previous and the current pilot subcarriers for the channel estimation at each received {\ac{OFDM}} symbol. On the contrary, in the proposed {\ac{WI}} estimators there are no fixed parameters, the time correlation between the previous and the future pilot symbols is considered in the {\ac{WI}} interpolation matrix~{\eqref{eq:ci}}, and the estimated channel at all channel taps is considered in the overall estimation. These aspects lead to the proposed estimators performance superiority.} 

In addition, {\ac{ChannelNet}} and {\ac{TS-ChannelNet}} estimators suffer from a considerable performance degradation that is dominant in very high mobility scenario. However, the proposed estimators show a robustness against high mobility, this is mainly due to the accuracy of the {\ac{WI}} interpolation, combined with optimized {\ac{SR-CNN}} and {\ac{DN-CNN}}. Although  {\ac{CNN}} processing is applied in the {\ac{ChannelNet}} and {\ac{TS-ChannelNet}}, this post {\ac{CNN}} processing is not able to perform well due to the high estimation error of the 2D RBF and ADD-TT interpolation techniques in the initial estimation. As a result, we can conclude that employing robust initial estimation as the proposed {\ac{WI}} interpolation schemes allows the {\ac{CNN}} to learn better the channel correlation with lower complexity, thus improving the channel estimation. Finally, we note that the performance of the 2D-LMMSE estimator is comparable to the performance of ideal channel but it requires huge complexity as we discuss in the next section, which is impractical in real scenario.
\begin{figure}[t]
  \centering
  \includegraphics[width=\columnwidth]{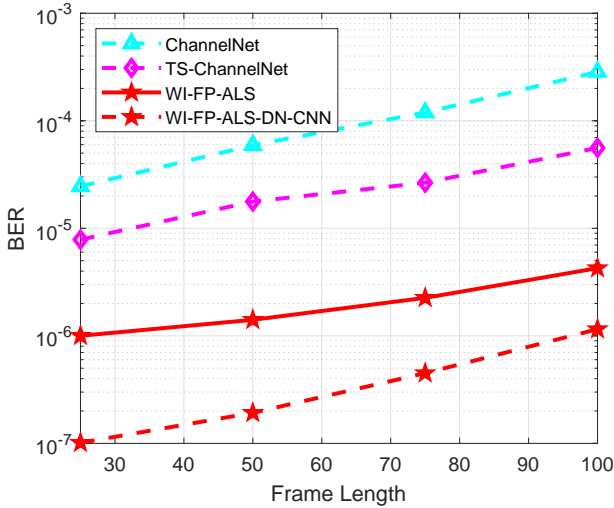}
  \caption{{BER performance of VTV-SDWW high mobility channel model employing QPSK and different frame lengths.}}
  \label{fig:ber_fl}
\end{figure}
\color{black}
\subsection{Frame Length}
{{\figref{fig:ber_fl}} shows the {\ac{BER}} performance of high mobility vehicular scenario employing QPSK modulation and different frame lengths. It can be clearly noticed that the proposed WI-FP-ALS estimator is able to outperform ChannelNet and TS-ChannelNet for different frame lengths, this is due to the robustness of the proposed WI-FP-ALS estimator, unlike the 2D RBF and ADD-TT interpolation techniques that suffer from a considerable estimation error even when a short frame is considered, which affects the performance of ChannelNet and TS-ChannelNet.  Moreover, employing the optimized DN-CNN after the WI-FP-ALS estimator improves significantly the BER performance.  {\tabref{tb:BER_HIGH_g}} illustrates the performance gain of the proposed WI-FP-ALS-DN-CNN estimator compared to the TS-ChannelNet estimator in high mobility scenario, where employing the optimized DN-CNN leads to  $5$ dB and $10$ dB gain in terms of SNR for a BER = $10^{-3}$ for QPSK and 16QAM modulation orders respectively as shown in {\figref{BER_16QAM_WI_CNN}}.}
\begin{table}[!t]
	\renewcommand{\arraystretch}{1.3}
	\small
	\centering
	\caption{BER performance gain (dB) of the proposed WI-FP-ALS compared to the TS-ChannelNet estimator in different mobility scenarios.}
	\label{tb:BER_HIGH_g}
	\begin{tabular}{|c|c|c|c|c|c|c|}
		\hline
		\multirow{2}{*}{\textbf{Scheme}} & \multicolumn{2}{c|}{\textbf{Low }} & \multicolumn{2}{c|}{\textbf{High }} & \multicolumn{2}{c|}{\textbf{Very High }} \\ \cline{2-7} 
		& WI                 & SRCNN                 & WI                  & DNCNN                 & WI                    & DNCNN                    \\ \hline
		\textbf{QPSK}                    & 4                  & 5                     & 3                   & 4                     & 2                     & 5                        \\ \hline
		\textbf{16QAM}                   & 2                  & 6                     & 3                   & 4                     & 5                     & 10                       \\ \hline
	\end{tabular}
\end{table}

\subsection{CNN Architecture}

The {\ac{ChannelNet}} estimator employs  {\ac{SR-CNN}} and {\ac{DN-CNN}} after the 2D RBF interpolation. The used {\ac{SR-CNN}} consists of three convolutional layers with $(v_{1} = 9 ; f_{1} = 64), (v_{2} = 1 , f_{2} = 32)$ and $(v_{3} = 5 , f_{3} = 1)$ respectively. Moreover, the {\ac{DN-CNN}} depth is $D = 18$  with $3 \times 3 \times 32$ kernels in each layer. On the other hand, {\ac{SR-ConvLSTM}} network consists of three ConvLSTM layers of $(v_{1} = 9 ; f_{1} = 64), (v_{2} = 1 , f_{2} = 32)$ and $(v_{3} = 5 , f_{3} = 1)$ respectively is integrated after the ADD-TT interpolation in the {\ac{TS-ChannelNet}} estimator. We note that, the {\ac{SR-ConvLSTM}} network combines both the {\ac{CNN}} and the LSTM networks~\cite{ref_ConvLSTM_Comp}, which  increases the overall computational complexity as we discuss later. In contrast, the employed optimized SR-CNN and DN-CNN decreases significantly the complexity due to the accuracy of the proposed WI estimators. In conclusion, as the accuracy of the pre-estimation increases, the complexity of the employed CNN decreases, since low-complexity architectures can be used and vice versa.  

%% file: complexity.tex
\section{Computational Complexity Analysis} \label{complexity}

In this section, a detailed computational complexity, {\ac{TDR}}, and latency analysis of the 2D LMMSE estimator, {\ac{ChannelNet}}, {\ac{TS-ChannelNet}}, and the proposed {\ac{WI}} estimators are presented.

\subsection{Computational Complexity Analysis}

The computational complexity is expressed in terms of real valued multiplication/division and summation/subtraction mathematical operations required for a full {\ac{OFDM}} frame channel estimation. Each complex-valued division requires $6$ real valued multiplications, $2$ real valued divisions, $2$ real valued summations, and $1$ real valued subtraction. On the other hand, each complex valued multiplication can be expressed by $4$ real valued multiplications, and $3$ real valued summations.

\subsubsection{2D LMMSE estimator} 

The conventional 2D LMMSE estimator requires first the {\ac{LS}} estimation as that requires $2 K_{p}I$ divisions. Then, the matrix inverse operation requires $4 K_{p}^{3} I^{3}$ multiplications and $3 K_{p}^{3} I^{3}$ summations. Finally, the correlation matrices are multiplied by the LS estimated channel vector resulting in $K_{p}^{2} I^{2} + K_{d}^{2} K_{p}^{2} I^{4} $ multiplications.
Therefore, the overall computational complexity of the conventional 2D LMMSE estimator is $4 K_{p}^{3} I^{3} + K_{p}^{2} I^{2} + K_{d}^{2} K_{p}^{2} I^{4} + 2 K_{p}I$ multiplications and $3 {K}_{p}^{3} I^{3} + 2 K_{p} I$ summations. We note that, in case the full $\ma{W}_{\text{2D-LMMSE}}$ matrix is calculated offline, the computational complexity of the 2D-LMMSE estimator is reduced to $4 K_{d} {K}_{p}^{2} I^{2} + 2 K_{p} I $ multiplications and $ 3 K_{d} {K}_{p}^{2} I^{2} + 2 K_{d} K_{p} I^{2} - 2 K_{d} I $ summations. We can notice that the 2D LMMSE suffers from very high computational complexity that make it impractical estimator in real-time scenarios.

\subsubsection{{\ac{ChannelNet}} estimator} employs the {\ac{RBF}} interpolation followed by {\ac{SR-CNN}} and {\ac{DN-CNN}} networks. Thereby, the computational complexity of the {\ac{ChannelNet}} is given by
\begin{equation}
\text{CC}_{{\text{ChannelNet}}} = \text{CC}_{{\text{RBF}}} + \text{CC}_{\text{SR-CNN}} + \text{CC}_{\text{DN-CNN}}.
\label{eq:CC_ChannelNet}
\end{equation}

As shown in~\eqref{eq: LS-RBF}, the calculation of $\hat{\tilde{\ma{H}}}_{\text{LS}}$ requires $2 K_{p}I$ divisions. $\ma{w}_{\text{RBF}}$ calculation requires $4 K^{2}_{p}I^{2} $ multiplications/divisions and $5 K^{2}_{p}I^{2} - 2 K_{p}I$ summations/subtractions. On the other hand, $\hat{\tilde{\ma{H}}}_{\text{RBF}}$ computation requires $K_{d}I (K^{2}_{p} I^{2} + 3K_{p}I)$ multiplications/divisions and $5K_{d}K_{p} I^{2}$ subtractions/summations. Therefore the total computational complexity of the {\ac{RBF}} interpolation can be expressed by $K^{2}_{p} I^{2} ( 4 + K_{d} I ) + K_{p} I (2 + 3 K_{d} I)$ multiplications/divisions and $ K_{p} I ( 5 K_{p} I + 5 K_{d}I - 2)$  summations/subtractions.
\begin{table*}[t]
\renewcommand{\arraystretch}{1.3}	
\centering
\caption{{CNN-based estimators Overall computation complexity in terms of the total required real valued operations.}}
\label{tb:comp}
\begin{tabular}{|c|c|c|c|c|}
\hline
\multirow{2}{*}{\textbf{Scheme}} & \multicolumn{2}{c|}{\textbf{Interpolation}} & \multicolumn{2}{c|}{\textbf{CNN}}       \\ \cline{2-5} 
                                 & \textbf{Mul./Div.}   & \textbf{Sum./Sub.}   & \textbf{Mul./Div.} & \textbf{Sum./Sub.} \\ \hline
ChannelNet  & $K^{2}_{p} I^{2} ( 4 + K_{d} I ) + K_{p} I (2 + 3 K_{d} I)$   & $ K_{p} I ( 5 K_{p} I + 5 K_{d}I - 2)$                      &   $350144 K_{\text{on}} I$                 &      $42432 K_{\text{on}} I$              \\ \hline
TS-ChannelNet    & $24 K_{\text{on}} I + 4 L K_{\text{on}} I$     &  $18 K_{\text{on}} I + 5 K_{\text{on}} I L$          &  $226880 K_{\text{on}} I$  &   $81472 K_{\text{on}} I$                 \\ \hline
FP-SLS-SR-CNN  &    $2 K_{\text{on}} P + 2 K_{\text{on}} + 4 K_{\text{on}} I_{d}$     &    $2 K_{\text{on}} + 2 K_{\text{on}} I_{d}$                      & \multirow{3}{*}{$7008 K_{\text{on}}I_{d}$}  & \multirow{3}{*}{$1120 K_{\text{on}}I_{d}$}  \\ \cline{1-3}
FP-ALS-SR-CNN   &   $4 K^{2}_{\text{on}} P + 2 K_{\text{on}} P + 2 K_{\text{on}} + 4 K_{\text{on}} I_{d}$   &    $5 K^{2}_{\text{on}} P + 2 K_{\text{on}} I_{d}$                     &                    &                    \\ \cline{1-3}
LP-SR-CNN    &         $2LP + 4 K_{\text{on}} L P + 2 K_{\text{on}} + 4 K_{\text{on}} I_{d} $ & $5 K_{\text{on}} L P + 2 K_{\text{on}} I_{d}$                        &                    &                    \\ \hline
FP-SLS-DN-CNN   &      $2 K_{\text{on}} P + 2 K_{\text{on}} + 4 K_{\text{on}} I_{d}$     &    $2 K_{\text{on}} + 2 K_{\text{on}} I_{d}$                      & \multirow{3}{*}{$84096 K_{\text{on}}I_{d}$}  & \multirow{3}{*}{$9856 K_{\text{on}}I_{d}$}  \\ \cline{1-3}
FP-ALS-DN-CNN   &      $4 K^{2}_{\text{on}} P + 2 K_{\text{on}} P + 2 K_{\text{on}} + 4 K_{\text{on}} I_{d}$   &    $5 K^{2}_{\text{on}} P + 2 K_{\text{on}} I_{d}$                  &                    &                    \\ \cline{1-3}
LP-DN-CNN    &   $2LP + 4 K_{\text{on}} L P + 2 K_{\text{on}} + 4 K_{\text{on}} I_{d} $ & $5 K_{\text{on}} L P + 2 K_{\text{on}} I_{d}$                       &                    &                    \\ \hline
\end{tabular}
\end{table*}
After that, the {\ac{ChannelNet}} estimator applies {\ac{SR-CNN}} followed by {\ac{DN-CNN}} on top of the {\ac{RBF}} interpolation. $\text{CC}_{\text{SR-CNN}}$ and $\text{CC}_{\text{DN-CNN}}$ can be computed as follows
\begin{equation}
\begin{split}
\text{CC}_{\text{SR-CNN}} &= \sum_{j=1}^{J} h_{j} w_{j} d_{j} v_{j}^2 f_{j}  + h_{j} w_{j} d_{j} f_{j} \\
&= \sum_{j=1}^{J} h_{j} w_{j} d_{j} f_{j} (v_{j}^2 + 1).
\end{split}
\label{CC_SR_CNN}
\end{equation}
\begin{equation}
\text{CC}_{\text{DN-CNN}} = \sum_{j=1}^{J}  h_{j} w_{j} d_{j} f_{j} (v_{j}^2 + 1) + \sum_{j=1}^{D} 4h_{j} w_{j} d_{j},
\label{CC_DN_CNN}
\end{equation}
where $J$ denotes the number of {\ac{CNN}} layers. We note that the second term in $\text{CC}_{\text{DN-CNN}}$ represents the number of operations required by the batch normalization considered in the {\ac{DN-CNN}} network. Therefore, the {\ac{SR-CNN}} used in the {\ac{ChannelNet}} estimator requires $16064 K_{\text{on}} I$ multiplications/divisions and $4288 K_{\text{on}} I$ summations/subtractions, while the {\ac{ChannelNet}} {\ac{DN-CNN}} computations require $334080 K_{\text{on}} I$ multiplications/divisions and $38144 K_{\text{on}} I$ summations/subtractions.

\subsubsection{{\ac{TS-ChannelNet}} estimator} applies the \ac{ADD-TT} interpolation followed by the  {\ac{SR-ConvLSTM}} network. Accordingly, the overall computational complexity of the {\ac{TS-ChannelNet}} estimator can be expressed as follows
\begin{equation}
\text{CC}_{{\text{TS-ChannelNet}}} = \text{CC}_{{\text{ADD-TT}}} + \text{CC}_{\text{SR-ConvLSTM}}.
\label{eq:CC_TS-ChannelNet}
\end{equation} 
The \ac{ADD-TT} interpolation applies two equalization steps~\eqref{eq: ADD_TT_1}, and~\eqref{eq: ADD_TT_2} after the {\ac{LS}} estimation in~\eqref{eq: LS1} that requires $2 K_{\text{on}}$ summations, and $2 K_{\text{on}}$ divisions. Each equalization step consists of $K_{\text{on}}$ complex valued division, therefore the equalization in~\eqref{eq: ADD_TT_1}, and~\eqref{eq: ADD_TT_2} requires  $16 K_{\text{on}}$ multiplications/divisions and $6 K_{\text{on}}$ summations/subtractions. The time domain truncation operation applied in~\eqref{eq:TT} requires $4 L K_{\text{on}}$ multiplications and $5K_{\text{on}} L - 2K_{\text{on}}$ summations. After time domain truncation, the {\ac{ADD-TT}} interpolation applies frequency and time domain averaging. The frequency domain averaging~\eqref{eq: STA_4} requires $10 K_{\text{on}}$ summations, and $2 K_{\text{on}}$ multiplications. Moreover, the time domain averaging step~\eqref{eq: STA_5}, requires $4 K_{\text{on}}$ real valued divisions, and $2 K_{\text{on}}$ real valued summations. Therefore, the overall computational complexity of the {\ac{ADD-TT}} interpolation for the whole received {\ac{OFDM}} frame requires $24 K_{\text{on}} I + 4 L K_{\text{on}} I$ real valued multiplications/divisions, and $18 K_{\text{on}} I + 5 K_{\text{on}} I L$ real valued summations/subtractions, and its computational complexity is expressed in terms of the overall operations applied in the input, forget, and output gates of the {\ac{SR-ConvLSTM}} network, such that

\begin{equation}
\text{CC}_{\text{ConvLSTM}} = \sum_{j=1}^{J}  h_{j} w_{j} d_{j} f_{j} (8v_{j}^2 + 30). 
\label{CC_ConvLSTM}
\end{equation}

Based on~\eqref{CC_ConvLSTM}, the {\ac{SR-ConvLSTM}} network utilized in the {\ac{TS-ChannelNet}} estimator requires $226880 K_{\text{on}} I$ multiplications/divisions and $81472 K_{\text{on}} I$ summations/subtractions.
{\ac{TS-ChannelNet}} estimator is less complex than the {\ac{ChannelNet}} estimator, since it employs only one {\ac{CNN}} on top of the {\ac{ADD-TT}} interpolation, unlike the {\ac{ChannelNet}} estimator where both {\ac{SR-CNN}} and {\ac{DN-CNN}} are used.
\subsubsection{Proposed {\ac{WI}} estimators} \ac{WI} estimators computational complexity depend mainly on the selected frame structure, the pilot allocation scheme, and the selected optimized {\ac{CNN}}. The overall computational complexity of the proposed {\ac{WI}} estimators can be expressed as follows
\begin{equation}
\text{CC}_{{\text{WI}}} = \text{CC}_{\hat{\tilde{\ma{H}}}_{\text{WI}}} + \text{CC}_{\text{O-CNN}}.
\label{eq:CC_WI}
\end{equation}


In case of inserting full pilot symbols, there are two options, {\ac{SLS}} estimator that requires only $2 K_{\text{on}} P + 2 K_{\text{on}}$ divisions, and $2 K_{\text{on}}$ summations. The second option is employing {\ac{ALS}}, where $2  K_{\text{on}} P + 2 K_{\text{on}}$ divisions, followed by  $4  K^{2}_{\text{on}} P$ multiplications, and $ 5 K^{2}_{\text{on}} P$ summations are required. On the other hand, when $K_{\text{p}} = L$ pilots are inserted with each pilot symbol, then the {\ac{LS}} estimation requires $2 L P + 2 K_{\text{on}} $ divisions, $4  K_{\text{on}} L P$ multiplications, and $ 5 K_{\text{on}} L P $ summations. After selecting the required frame structure and pilot allocation scheme, the proposed estimators apply the weighted interpolation as shown in~\eqref{eq:WI_LS}, where the channel estimation for each received {\ac{OFDM}} frame requires $4 K_{\text{on}}I_{d}$ divisions and  $2 K_{\text{on}}I_{d}$ summations. Finally, the optimized {\ac{SR-CNN}} is utilized in low mobility scenario and it requires $7008 K_{\text{on}}I_{d}$ multiplications/divisions and  $1120 K_{\text{on}}I_{d}$ summations/subtractions, while the optimized {\ac{DN-CNN}} is exploited in high and very high mobility scenarios requires $84096 K_{\text{on}}I_{d}$ multiplications/divisions and $9856 K_{\text{on}}I_{d}$ summations/subtractions

\begin{figure}[t]
\centering
\begin{tikzpicture}
\begin{axis}[
    ybar,
    ylabel={Real Valued Operations x 1000},
    symbolic x coords={FP-ALS, LP, FP-SLS},
    xtick=data,
    legend style={at={(0.48,+1.2)},
    anchor=north,legend columns=-1},
    nodes near coords align={vertical},
    width=\columnwidth,
    height=6cm,
    grid=major,
    cycle list = {black,black!70,black!40,black!10}
    ]
\addplot+[] coordinates {(FP-ALS, 42.640) (LP, 25.528) (FP-SLS, 20.696 )};
\addplot+[fill,text=black!10] coordinates {(FP-ALS,  37.232) (LP, 16.432) (FP-SLS, 10.296) };
\legend{Multiplications/Divisions, Summations/Subtractions}
\end{axis}
\end{tikzpicture}
\caption{Computational complexities of the proposed WI estimators employing P = 2.}
\label{fig:bar_graph_WI}
\end{figure}
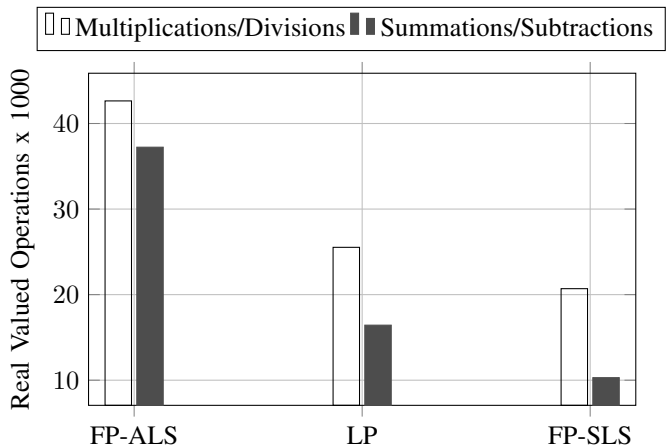
The proposed {\ac{WI}}-FP-ALS records the higher computational complexity among the other proposed estimators in all mobility scenarios, due to the $\ma{W}_{\text{ALS}}$ calculation in~\eqref{eq:ALSP}. Moreover, the proposed {\ac{WI}}-FP-SLS estimator is the simplest one. \figref{fig:bar_graph_WI} shows the computational complexity of the proposed {\ac{WI}} estimators employing $P = 2$ pilot symbols within the frame.

\tabref{tb:comp} shows the overall computational complexity of the benchmarked estimators in terms of real valued operations. It is worth mentioning that the proposed {\ac{WI}} estimators achieve a significant computational complexities decrease compared to {\ac{ChannelNet}} and {\ac{TS-ChannelNet}} estimators, where
 {\ac{ChannelNet}} and {\ac{TS-ChannelNet}} estimators are $70$ and $39$ times more complex than the proposed FP-ALS-SR-CNN estimator respectively. Moreover, the proposed estimators achieve at least $7027.35$ times less complexity than the 2D LMMSE estimator, with an acceptable {\ac{BER}} performance, thus making them a good alternative to the 2D LMMSE. We note that FP-ALS-DN-CNN is $12$ times more complex than FP-ALS-SR-CNN since the optimized {\ac{DN-CNN}} architecture complexity that is employed in high and very high scenarios is higher than the optimized {\ac{SR-CNN}} architecture which is used in low mobility scenario.


\subsection{Transmission data rate and latency analysis}

The \ac{TDR} and the latency introduced at the receiver in order to recover the transmitted bits are important issues in vehicular communications, especially for real time applications~\cite{ref_latency}. The transmission data rate is influenced by the number of allocated data subcarriers within the transmitted frame:
\begin{equation}
\text{TDR} = \frac{K_{\text{DF}}\log_{2} (M) \rho }{T_{\text{s}} I},
\end{equation}
where $K_{\text{DF}}$ and $\rho$ denote the total allocated data subcarriers within the transmitted frame, and the employed code rate respectively, and $M$ represents the utilized modulation order. Moreover, the buffering time $\varphi$ can be expressed by the total duration that the receiver should wait before starting the channel estimation, such that
$\varphi = T_{\text{s}} I$
,where $T_{\text{s}}$ represents the received {\ac{OFDM}} data symbol duration.
\ac{ChannelNet} and {\ac{TS-ChannelNet}} estimators employ $K_{p} = 4$ pilot subcarriers within each transmitted {\ac{OFDM}} symbol, however \ac{ChannelNet} estimator requires sparsed pilot allocation within the transmitted {\ac{OFDM}} frame, and thereby, increasing the complexity of pilots extraction at the receiver, since the allocated pilots subcarriers indices differs between the {\ac{OFDM}} symbols within the frame. Therefore, {\ac{ChannelNet}} and {\ac{TS-ChannelNet}} estimators have similar transmission data rate as defined in the IEEE 802.11p standard. However, they suffer from high buffering time  at the receiver, since the full frame should be received before the channel estimation starts leading to high latency.

\begin{figure}[t]
  \includegraphics[width=\columnwidth]{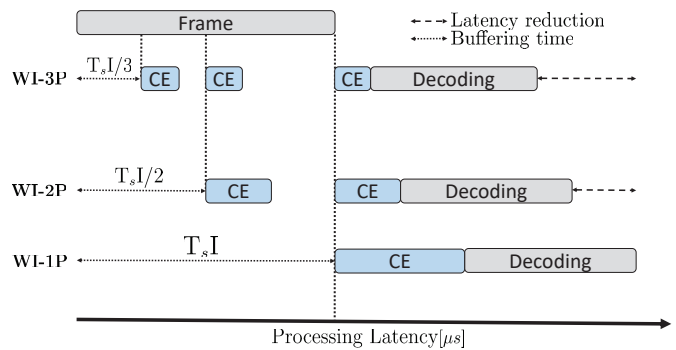}
  \caption{Processing latency of the proposed {\ac{WI}} estimators.}
  \label{fig:block_latency}
\end{figure}

\begin{table*}[t]
\renewcommand{\arraystretch}{1.3}
\small
\centering
\caption{Transmission data rate and buffering time analysis for the benchmarked schemes.}
\label{tb:TDR_Analysis}
\begin{tabular}{|c|c|c|c|c|c|c|c|c|}
\hline
\multirow{3}{*}{\textbf{Scheme}} & \multirow{3}{*}{\textbf{ChannelNet}} & \multirow{3}{*}{\textbf{TS-ChannelNet}} & \multicolumn{6}{c|}{\textbf{Proposed 2D WI}}     \\ \cline{4-9}  & &   & \multicolumn{2}{c|}{WI-1P} & \multicolumn{2}{c|}{WI-2P} & \multicolumn{2}{c|}{WI-3P} \\ \cline{4-9} 
& & & FP & LP & FP & LP & FP & LP    \\ \hline
\textbf{TDR gain} & 0 \% & 0\%& 7.25\% &  8.08\%  & 6.16\% & 7.83\% & 5.08\%  & 7.58\%   \\ \hline
\textbf{$\varphi$ [$\mu$s]} & 800 & 800 & \multicolumn{2}{c|}{800} & \multicolumn{2}{c|}{400} & \multicolumn{2}{c|}{265}         \\ \hline
\end{tabular}
\end{table*}

As shown in \tabref{tb:TDR_Analysis}, the proposed {\ac{WI}} estimators record different {\acp{TDR}} gains according to the selected frame structure. Moreover, the proposed {\ac{WI}}-2P and {\ac{WI}}-3P  estimators require lower buffering time than the proposed {\ac{WI}}-1P, {\ac{ChannelNet}}, and {\ac{TS-ChannelNet}} estimators, since it divides the frame into several sub frames as shown in \figref{fig:block_latency}. Hence, the channel estimation process starts earlier. Therefore, the proposed {\ac{WI}} estimators contribute in reducing the total required latency. Finally, we note that the transmission parameters and the chosen frame structure should be adapted according to the mobility condition, required data rate, and the acceptable latency by the vehicular application.

%% file: conclusions.tex
\acresetall
\section{Conclusion} \label{conclusions}
In this paper, FBF channel estimation in vehicular communication is studied, where the limitations of the conventional 2D LMMSE estimator and the motivation behind employing CNN processing in the channel estimation are presented. Moreover, The recently proposed CNN-based channel estimators have been extensively surveyed. In this context, we have proposed a hybrid, adaptive, and robust WI channel estimators for the IEEE 802.11p standard, where pilot symbols are inserted within the transmitted frame, with several pilot allocation schemes adapted according to the mobility condition. Unlike the recently proposed ChannelNet and TS-ChannelNet estimators that suffer from high computational complexity, performance degradation in high mobility vehicular scenarios, and high latency at the receiver, the proposed WI estimators have reduced computational complexity and robustness in high mobility scenarios. Moreover, they require low buffering time at the receiver and more TDR gain is achieved since all the OFDM symbols within the transmitted frame are fully allocated to data. Additionally, the employed SR-CNN and DN-CNN architectures are optimized through intensive experiments in order to alleviate the high complexity problem. Simulation results have shown the performance superiority of the proposed WI estimators over ChannelNet and TS-ChannelNet estimators in all vehicular scenarios with a substantial reduction in computational complexity, where ChannelNet and TS-ChannelNet are more complex than the proposed WI-FP-ALS-SR-CNN by $70$ and $39$ times respectively. On the other hand, the proposed estimators are less complex than the conventional 2D LMMSE estimator by at least $7027.35$ times while recording a convenient BER performance especially in high mobility vehicular scenarios, which makes them good alternatives to the conventional 2D LMMSE estimator.